\documentclass[twocolumn,twocolappendix]{aastex631}

\usepackage{amsmath}

\newcommand{\Msun}{\mathrm{M}_\odot}

\newcommand{\pkg}[1]{\textit{#1}}

\newcommand{\cGpc}{\mathrm{cGpc}}

\newcommand{\kpc}{\mathrm{kpc}}

\newcommand{\cMpc}{\mathrm{cMpc}}

\newcommand{\ckpc}{\mathrm{ckpc}}

\newcommand{\Msunyr}{\Msun\, \mathrm{yr}^{-1}}
\newcommand{\Mstar}{M_\mathrm{*}}
\newcommand{\Mgas}{M_\mathrm{gas}}
\newcommand{\Mvir}{M_\mathrm{vir}}
\newcommand{\Rvir}{R_\mathrm{vir}}

\newcommand{\Omegamnow}{\Omega_\mathrm{m,0}}
\newcommand{\Omegalambnow}{\Omega_\mathrm{\Lambda,0}}
\newcommand{\Gyr}{\mathrm{Gyr}}
\newcommand{\Myr}{\mathrm{Myr}}
\newcommand{\yr}{\mathrm{yr}}
\newcommand{\sigmaeight}{\sigma_\mathrm{8}}

\newcommand{\masstwo}{M_\mathrm{200,c}}
\newcommand{\massfive}{M_\mathrm{500,c}}

\newcommand{\rhocrit}{\rho_\mathrm{crit}}

\newcommand{\rtwo}{R_\mathrm{200,c}}
\newcommand{\rfive}{R_\mathrm{500,c}}

\newcommand{\rnaught}{r_\mathrm{0}}

\newcommand{\Reff}{R_\mathrm{eff,*}}
\newcommand{\Mbhmstar}{M_\mathrm{BH} / M_\mathrm{*}}
\newcommand{\ergs}{\mathrm{erg}\,\mathrm{s}^{-1}}

\begin{document}

\title{The Manhattan Suite: Accelerated galaxy evolution in the early Universe}

\correspondingauthor{Douglas Rennehan}
\email{douglas.rennehan@gmail.com}

\author[0000-0002-1619-8555]{Douglas Rennehan}
\affiliation{Center for Computational Astrophysics,
Flatiron Institute, 162 5th Ave, New York, NY 10010, USA}



\begin{abstract}

Observational advances have allowed the detection of galaxies, protoclusters, and galaxy clusters at higher and higher redshifts, opening a new view into extreme galaxy evolution.  I present an argument that the high redshift, massive galaxies discovered over the last decade are really the most massive galaxies within protocluster-cores of galaxy clusters at $z\sim2$, and that they are the partial descendants of same galaxies discovered by JWST at $z\sim9$. To that end, I present \textit{The Manhattan Suite}, a set of $100$ high resolution zoom-in simulations of the most massive galaxy clusters, out to $9\,\Rvir$, selected at $z = 2$ from a ($1.5\,\cGpc)^3$ parent volume, and simulated using the \pkg{Simba} model.  Unlike other cluster suites, my selection at $z = 2$ ensures that these systems are biased in a similar fashion to observations, in that they should be the brightest and the most massive by construction at $z \gtrsim 2$.    I show that my sample is able to reproduce extremely star-bursting protoclusters such as SPT2349-56, high redshift galaxy clusters XLSSC122 and JKCS041, and the wealth of massive (sometimes quenched) galaxies at $z \gtrsim 3$ and up to $z \sim 9$.  I argue that these systems are intimately linked, and represent the same evolutionary history. 
\end{abstract}



\section{Unraveling rarity} \label{sec:intro} 

Under the $\Lambda\mathrm{CDM}$-paradigm, the first galaxies in the Universe form in the gravitational depths dark matter halos, where gas cools and coalesces to form stars \citep{White1978}.  Slowly over cosmic time, smaller dark matter halos merge hierarchically under the influence of gravity, leading to more and more massive structures.  Galaxy clusters are at the peak of the hierarchy, and contain thousands of individual galaxies within a single gravitationally-bound structure \citep{Bahcall1977}.  However, observations of mature galaxy clusters at $z \sim 2$ complicate the picture as they only have roughly $\sim4\,\Gyr$ to undergo all of their evolution \citep{Andreon2009, Andreon2014, Willis2020, Andreon2021}, which was once thought to occur over the entire history of the Universe (see \citealt{Kravtsov2012} for a review).  Additionally, the discovery massive ($\gtrsim10^{11}\,\Msun$) galaxies \citep{Glazebrook2017, Valentino2020, Xiao2023, deGraff2024, Jin2024, Kakimoto2024} and extremely star forming protoclusters ($\sim10,000\,\Msunyr$) \citep{Ishigaki2015, Jiang2018, Miller2018, Higuchi2019, Ito2019, Hill2022, AlvarezMarquez2023} at high redshift ($z \gtrsim 2$) have challenged our theoretical models of galaxy evolution.  

Over the past decade there has been an unrelenting quest for the discovery of galaxies at higher and higher redshifts.  Before the launch of JWST, the majority of spectroscopically confirmed galaxies were determined to exist roughly at $z\sim7$ after the Big Bang (e.g. \citealt{Finkelstein2013, Oesch2015, Hasimoto2018}).  However, that quickly changed as a flurry of new observations from JWST broadened the picture even further, as the resolution and sensitivity of the instrument provides a new insight into the early Universe \citep{JWST2023}.  In particular, NIRCam photometry \citep{NIRCam2023} allowed the construction of luminosity functions as high as $z \sim 11$ (in \pkg{CEERS}; \citealt{Finkelstein2023}) and multiple confirmed detections of galaxies at $z \sim 13$ \citep{CurtisLake2023, Robertson2023}.

Not only were galaxies found earlier than expected, there were also photometrically-identified, overly massive galaxies at ultra-high redshifts ($\Mstar\sim10^{10}\,\Msun$ at $z\gtrsim7$; \citealt{Labbe2023}),  that were seemingly in stark disagreement with $\Lambda\mathrm{CDM}$ predictions \citep{BoylanKolchin2023}.  However, spectroscopic follow-up of these sources showed that (a) their redshifts were either much lower ($z\sim5$) than identified by photometry alone, (b) that the stellar masses were much smaller, or (c) that they were obscured active galactic nuclei (AGN) \cite{Endsley2023, Fujimoto2023, Kocevski2023} --- matching the expectations from pre-JWST examples of overly massive galaxies (see e.g, \citealt{Glazebrook2017, Valentino2020}).  There are several examples of spectroscopically confirmed massive galaxies at high redshift from recent JWST surveys such as \pkg{RUBIES} (JWST-GO-4233; PI de Graaff; $\Mstar\sim10^{11}\,\Msun$ at $z\sim5$ from \citealt{deGraff2024}), \pkg{UNCOVER} (\citealt{Bezanson2022}; $\Mstar\sim2\times10^{10}\,\Msun$ at $z\sim4$ from \citealt{Setton2024}), and \pkg{COSMOS-Web} (\citealt{Casey2023}; $\Mstar\sim10^{12}$ from \citealt{Lambrides2024}, photometric redshift $z\sim7$).  The study in \cite{Xiao2023} using the JWST \pkg{FRESCO} survey \citep{Oesch2023} in the \pkg{GOODS} North and South fields unveiled three extremely massive galaxies with $\Mstar \gtrsim 10^{11}\,\Msun$ all spectroscopically confirmed above $z > 5$.  The issue with these particular observations is that not only are these galaxies extremely massive at early times, but they are already quenched with specific star formation rates (sSFRs) of $sSFR \lesssim10^{10}\,\mathrm{yr}^{-1}$, indicating that they have already gone through their entire evolution early in the Universe (recall the observed $z\sim2$ clusters above).  It is possible that these galaxies have (a) consumed all of their available gas \citep{Kimmig2024}, or (b) have been quenched due to AGN feedback \citep{Spzila2024, Xie2024}. Regardless of the quenching mechanism, the galaxies must have had substantial star formation rates (SFRs) in the past in order to grow so quickly \citep{deGraff2024}.

It seems reasonable to expect that the high redshift quenched galaxies may have begun as highly starbursting galaxies within protocluster cores. Indeed, there have been observations of more and more extreme protocluster regions, forming stars at rates of $1000\,\Msunyr$ to $10,000\,\Msunyr$ within their cores --- the region that should eventually form a brightest cluster galaxy (BCG).   One particular protocluster, SPT2349-56 \citep{Miller2018}, is observed at $z\sim4$ and contains $14$ massive ($\Mstar\sim10^{10}\,\Msun$) galaxies all within a region $130\,\kpc$ (physical) on the sky, and has an estimated SFR of $\sim5,000\,\Msunyr$ within the observed core of the structure, and $\sim10,000\,\Msunyr$ in the extended region \citep{Hill2022}.  We showed in \cite{Rennehan2020} that a BCG should form within $\sim300\,\Myr$ and reach an SFR of $\sim3000\,\Msunyr$, by constructing a bespoke simulation of the SPT2349-56 structure given the observations in \cite{Miller2018}.   However, large-scale cosmological simulations have been unable to reproduce the properties such as the low depletion timescales, high gas fractions, and the high sSFR of the SPT2349-56 system \citep{Yajima2021, Remus2023}, challenging the robustness of our galaxy evolution models (see e.g. \citealt{Lim2020}).  The protocluster regions at high redshift already contain enormous stellar masses ($10^{12}-10^{13}\,\Msun$) \citep{Hill2022}, indicating that a subset of galaxies must form rapidly at $z \gtrsim 6$.

There are observed examples of BCGs that have seemingly formed $10^{12}\,\Msun$ of stars by $z \sim 1.5$ \citep{Collins2009}, leaving no room for further growth.  The picture is further complicated due to there being little evidence for size evolution in the BCG population after $z\sim1$, indicating that most of the mass should grow before that time \citep{Whiley2008, Stott2011}.  Furthermore, an analysis of the Spitzer Adaptation of the Red-Sequence Cluster Survey (SpARCS; \citealt{Muzzin2009, Wilson2009}) indicates that the SFRs within a subset of BCGs \textit{increases} as a function of redshift, reaching values as high as $\sim1000\,\Msunyr$ by $z\sim1.5$ \citep{Webb2015}.  These observations suggest that there is a distribution of growth histories of BCGs, and that the progenitors of a subset of these galaxies should be proto-BCGs that are rapidly forming stars within protocluster regions at $z \gtrsim 2$ \citep{Rennehan2020}.  It is natural to assume that a subset of these BCGs could be forming within regions that would become galaxy clusters at $z\sim2$, such as the observed XLSSC122 \citep{Willis2020} and JKCS041 \citep{Andreon2009} systems.

While there are physical explanations for how these galaxies may quench, the estimated high abundances ($\sim10^{-6}\,\cMpc^{-3}$) of the massive quenched galaxies suggests that our theoretical galaxy evolution and cosmological models may be lacking \citep{deGraff2024}.  It is important to interpret the JWST results in terms cosmic variance, as the field-to-field variance is substantial in small fields \citep{Bhowmick2020}, and can change the interpreted abundances (see e.g., \citealt{Moster2011, Yung2022a, Yung2022b}).  The impact of cosmic variance is not only significant in small fields, but also when galaxies are strongly clustered --- a particular problem for massive galaxies \citep{Somerville2004}.  

Our primary tool to interpret the observations -- cosmological simulations -- suffer from a different kind of variance, where the volume of the box plays an important role \citep{dePutter2012}.  Initial conditions are generally constructed by sampling the $\Lambda\mathrm{CDM}$ power spectrum modes into a finite cubical or spherical volume $V\propto L^3$ \citep{Hahn2011}.  However, since the volume is finite, only modes smaller than the box size $L$ are sampled.  Simultaneously, the volumes are periodic which leads to \textit{ultra-mean} regions of space.  As $L$ increases, rarer and rarer objects begin to appear, leading to an increase in the variance of overdensities represented.  For example, a $L \sim 150\,\cMpc$ volume would roughly give you a single object with number density $n\sim3\times10^{-7}\,\cMpc^{-3}$ (e.g. a $10^{15}\,\Msun$ galaxy cluster), whereas you would need $L\sim750\,\cMpc$ to capture $100$ objects for statistics.  It is generally computationally prohibitive to simulate these regions with hydrodynamics and gravity, yet certain groups have been able to push the boundaries with $L = 950\,\cMpc$ (Magneticum; Dolag et al., in prep.), $L = 740\,\cMpc$ (MilleniumTNG; \citealt{Pakmor2023}) and $L = 1000 \,\cMpc$ (FLAMINGO; \citealt{Schaye2023}) volumes at reasonable resolutions, with full galaxy evolution models.  

There are other methods to improve variance such as the zoom-in simulation technique \citep{Katz1993}, which allows sub-regions sampled from large volumes to be simulated at a much higher resolution.  Usually the approach targets halos of a certain mass or formation history and has been used in attempt to understand high redshift protoclusters in small samples, at fixed halo mass \citep{Trebitsch2021, Yajima2021}.  A novel method used in the \pkg{FLARES} project is to simulate a small sample of sub-regions,  roughly $L = 22\,\cMpc$, that statistically sample the underlying galaxy number density distribution from very large volumes, $L = 3200 \,\cMpc$ \citep{Lovell2020}.  The technique has been extremely successful in explaining the observed JWST mass functions at high redshift, indicating that capturing the variance of the galaxy overdensity distribution is necessary \citep{Keller2023}, and from extremely large volumes. 

In this paper, my goal is to statistically sample rare objects with mass $\Mvir\sim10^{14}\,\Msun$ at $z = 2$, which should represent the most rapid halo assembly -- and, therefore, galaxy assembly -- in the early Universe.  Therefore, I introduce a set of $100$ zoom-in simulations of galaxy clusters (and protocluster regions) selected as the $100$ most massive halos from a $(1.5\,\cGpc)^3$ dark matter only simulation.   In Section~\ref{sec:methods}, I overview my sample selection and the \pkg{Simba} galaxy evolution model that I use in this work.  In Section~\ref{sec:galaxy_populations}, I discuss how the suite reproduces the properties of the observed $z \sim 2$ mature galaxy clusters XLSSC122 and JKCS041, the extreme SPT2349-56 star-bursting protocluster, as well a range of observed massive galaxies from $2 \lesssim z \lesssim 9$.  In Section~\ref{sec:galaxy_monsters}, I compare the progenitor histories of the most massive galaxies in my protocluster sample to observed high-redshift galaxies. In Section~\ref{sec:discussion}, I discuss the connection between these objects and the implications of sampling the correct overdensity variance.  In Section~\ref{sec:conclusions}, I draw conclusions.

\section{Methodology} \label{sec:methods}

\subsection{Cosmology} \label{sec:methods_cosmology}

In this work, I assume a cosmology consistent with the \cite{Ade2016} results, having $\Omegamnow = 0.308$, $\Omegalambnow = 0.692$, $h = 0.6781$, $\Omega_\mathrm{b,0} = 0.0484$, $n_\mathrm{spec} = 0.9677$, and $\sigmaeight = 0.8149$.  Whenever I use the virial mass $\Mvir$ and virial radius $\Rvir$, I use the overdensity threshold, $\Delta_\mathrm{c} = 18\pi^2 + 82(\Omega_\mathrm{m}(z) - 1) - 38(\Omega_\mathrm{m}(z) - 1)^2$, where $\Omega_\mathrm{m}(z) \equiv \Omegamnow (1 + z)^3 / H(z)^2$ and $H(z) \equiv \sqrt{\Omegamnow(1 + z)^3 + \Omegalambnow}$ \citep{Bryan1998}.  Specifically, $\Mvir$ is defined such that the average density enclosed within $\Rvir$ is $\rho_\mathrm{avg} = \Delta_\mathrm{c} \rhocrit$, where $\rhocrit(z) \equiv 3H(z)^2/(8\pi G)$ and $G$ is Newton's gravitational constant.  In the case of $\masstwo$ and $\rtwo$, I use $\Delta_\mathrm{c} = 200$. Likewise, in the case of $\massfive$ and $\rfive$ which are common when analyzing galaxy clusters, I use $\Delta_\mathrm{c} = 500$.

\subsection{Initial conditions} \label{sec:methods_ics}

Given the discovery of massive, relaxed clusters at $z \sim 2$ (e.g. \citealt{Andreon2009, Willis2020}), I choose to make a selection of galaxy clusters at $z = 2$, using the definition in $\Mvir > 10^{14}\,\Msun$.  These clusters should represent the densest regions in the early Universe and, therefore, should host the most highly star forming galaxies at high redshift \citep{Chiang2017}.  Due to computational expense, I will limit the sample to $100$ of the most massive galaxy clusters at $z = 2$ selected from a large-volume dark matter only simulation. In particular, I use the \pkg{Triple Whopper} simulation to select the resimulation regions.

The \pkg{Triple Whopper} simulation is a $2048^3$ dark matter only, N-body simulation in a volume with side length $L = 1.5 \, \cGpc$ and is part of the \pkg{Burger Suite} of simulations (Steinwandel \& Rennehan, in prep.).   The simulation was run with \pkg{Gadget-4} \citep{Springel2021} and has a mass resolution $M \approx 1.5\times10^{10}\,\Msun$ and spatial resolution $\epsilon_\mathrm{g} = 18 \, \ckpc$.  The mass resolution allows me to pick halos of mass $M_\mathrm{vir} \approx 1.5\times10^{13}\,\Msun$ reliably, with about $1000$ particles \citep{Onorbe2013}.  The simulation uses the same cosmology that I outline in Section~\ref{sec:methods_cosmology}, and the initial redshift of the simulation is $z = 124$. 

To find the most massive structures at $z = 2$, I use the \pkg{Rockstar} package\footnote{\url{https://bitbucket.org/gfcstanford/rockstar}} \citep{Behroozi2013}.  \pkg{Rockstar} is a hierarchical friends-of-friends (FoF) finder that uses the full 6D configuration space, in addition to time, to find structures in a simulation and grid independent fashion.  Once I have a halo catalogue at $z = 2$, I select the most massive $100$ clusters and select all of the unique particle identifiers that are within a sphere of radius $R = 9 \,\Rvir$ from the halo minimum potential center.  I trace these particles back in time to the initial condition of the simulation, and find their initial positions.  I use the \pkg{MUSIC}\footnote{\url{https://www-n.oca.eu/ohahn/MUSIC/}} software \citep{Hahn2011} and these positions to generate new initial conditions by adding shorter wavelength modes and baryons in the high resolution region, all while degrading the resolution of the parent simulation volume.  After this process, the effective resolution in the zoom region for each halo is $8192^3$, which corresponds to a mass resolution of $m_\mathrm{dark} = 1.9\times10^8\,\Msun$ for dark matter and $m_\mathrm{gas} \approx 3.5\times10^7\,\Msun$ for gas particles.

\subsection{Code and Hydrodynamics solver} \label{sec:methods_code_hydro}

For the initial set of runs, I choose to use the \pkg{Simba} galaxy formation model as presented in \cite{Dave2019}.  The \pkg{Simba} model has been successful in reproducing the broad galaxy population, and especially the high redshift star forming galaxy population \citep{Lovell2021}.  It is based on \pkg{MUFASA} \citep{Dave2016c}, and includes models for cooling, star formation, stellar feedback, and black hole evolution and feedback, all while using the mesh-free finite mass (MFM) hydrodynamics method \citep{Lanson2008a, Lanson2008b, Gaburov2011}.  The model is built into \pkg{GIZMO} \citep{Hopkins2015a}, a publicly available hydrodynamics+gravity solver.  GIZMO is equipped with many features, including adaptive gravitational softening \citep{Price2007, Hopkins2017} which, at simulation time, computes the ideal gravitational softening length $\epsilon_\mathrm{i,grav}(t)$ for each particle $i$.  However, it requires a minimum value $\epsilon_\mathrm{grav,min}$ to prevent runaway, low values. I follow \cite{Dave2019} and use a minimum value of $\epsilon_\mathrm{grav,min} = 0.73\,\ckpc$.   Below, I briefly describe the simulation and sub-grid galaxy formation models, but point the reader to \cite{Hopkins2015a, Hopkins2016b}, and \cite{Dave2019} for a more complete description.

\subsection{Cooling and star formation}
\label{sec:methods_cooling_star_formation}

\pkg{Simba} uses the the Grackle chemistry and cooling library\footnote{\url{https://grackle.readthedocs.io/}} \citep{Smith2017} to compute primordial and radiative cooling in the presence of metals, as well as photo-heating and photo-ionization from the ultraviolet background \citep{Faucher2009} -- including self-shielding \citep{Haardt2012}.  To account for metal line cooling, \pkg{Simba} tracks 11 elements (H, He, C, N, O, Ne, Mg, Si, S, Ca, and Fe) which account for most of the mass in the Universe. 

In the model, once gas cools and becomes sufficiently dense it may form stars.  \pkg{Simba} uses the fraction of H$_\mathrm{2}$-gas to compute the SFR in a single gas element.  The H$_\mathrm{2}$ rate is computed via the model of \cite{Krumholz2009} and depends on the local metallicity and column density of the gas.  Specifically, the SFR is set by

\begin{equation}
    \label{eq:h2_star_formation_rate}
    \dot{\rho}_\mathrm{sfr} \equiv \frac{\epsilon_\mathrm{*}\rho_{\mathrm{H}_\mathrm{2}}}{t_\mathrm{dyn}},
\end{equation}

\noindent where $\epsilon_* = 0.02$ is the star formation efficiency, $t_\mathrm{dyn}$ is the dynamical time, and $\rho_{\mathrm{H}_\mathrm{2}}$ is the density of molecular gas within the gas particle.  

Once the rate of star formation, $\dot{\rho}_\mathrm{sfr}$, is known, stars stochastically form from the gas particles to satisfy the expected stellar mass formed in a single time step.  In particular, gas particles are converted directly into individual massive star particles that represent entire stellar populations with mass and metallicity inherited from their parent gas particle.

\subsection{Stellar feedback}
\label{sec:methods_stellar_feedback}

Over the course of their lives, stars eject mass, metals, and energy into their environments through stellar winds and supernova explosions.  These processes occur below the resolution scale of our simulations and, therefore, I require \textit{sub-grid} models for these physical processes.  In the following discussion, star particles are stellar populations with total mass $M$, metallicity $Z$, and age $t_\mathrm{age}$, and are distributed with a \cite{Chabrier2003}  initial mass function.

Massive stars in a stellar population are short-lived (relative to galactic time scales) and eject massive amounts of energy ($\sim 10^{51} \, \mathrm{erg}$) through the supernova type II (SNII) process.  The amalgamation of SNII across an entire galaxy leads to a large-scale galactic wind that can impact the gas in, and surrounding, the galaxy.  The \pkg{Simba} model involves launching a stochastic, kinetically driven wind from star forming gas particles.  The model has two parameters: the mass loading $\eta = \dot{M}_\mathrm{wind} / \dot{M}_*$ and the wind speed $v_\mathrm{wind}$.  Note that the mass loading directly sets the mass rate expected in winds per timestep, $\dot{M}_\mathrm{wind} = \eta\dot{M}_*$, for each star forming gas particle.

The mass loading comes from an analysis of the \pkg{FIRE} zoom-in simulations \citep{Hopkins2014} that tracked individual gas particles out through $25\%$ of the virial radius \citep{Angles2017}.  This procedure was done for galaxies of various stellar masses, and the mass loading was found to be a broken power law.  The wind velocities follow from another analysis of the \pkg{FIRE} simulations in \cite{Muratov2015}, but with a tunable amplitude. The \pkg{Simba} model also accounts for the gravitational potential difference between where the wind is launched (near the galaxy) and $25\%$ of the virial radius ($R_\mathrm{vir}$).  

The \pkg{Simba} model uses a two component model for supernova type Ia (SNIa) with a prompt (coincident with star formation) and delayed component, following the rates in \cite{Scannapieco2005}.  The energy form the prompt component ($10^{51}\,\mathrm{erg}$) directly heats and enriches the star forming gas particle, whereas the delayed component occurs $0.7$ $\Gyr$ after the star formation time and contributes to the nearest $16$ gas particle neighbors.

Asymptotic giant branch (AGB) stars also provide a feedback channel in the \pkg{Simba} model.  Specifically, \pkg{Simba} uses the model of \cite{Conroy2015a} and compute a stellar mass-loss rate based on the work of \cite{Bruzual2003}. AGB feedback is delayed substantially relative to the initial formation of the stellar population, and the energy rate computed from the mass loss rate is deposited directly into the $16$ nearest gas particle neighbors.  Additionally, those particles are pushed at a velocity of $100$ $\mathrm{km}$ $\mathrm{s}^{-1}$ radially from the stellar population.

\subsection{Chemical enrichment and dust}
\label{sec:methods_chemical_enrichment_and_dust}

SNII, SNIa, and AGB stars all contribute metals in their environments, eventually leaving interstellar medium and enrich regions far beyond the galaxy.  As I mentioned in Section~\ref{sec:methods_cooling_star_formation}, \pkg{Simba} tracks H, He, C, N, O, Ne, Mg, Si, S, Ca, and Fe that represent the majority of metal mass in the Universe\footnote{See \citealt{Hough2023} for an updated chemical enrichment model that tracks $34$ elements.}.  \pkg{Simba} uses SNII yields from \cite{Nomoto2006}, SNIa yields from \cite{Iwamoto1999}, and yields for AGB stars as described in \cite{Oppenheimer2008}.

\pkg{Simba} has a model for the formation, evolution, and destruction of dust produced in the ejecta of supernova explosions and AGB stars.  Dust directly follows gas particles in the simulation and is represented as a fractional mass of the gas particles, similar to the metallicity.

\subsection{Black hole physics}
\label{sec:methods_black_hole_physics}

\subsubsection{Seeding and dynamics}
\label{sec:methods_black_hole_seeding_and_dynamics}

Black holes are first seeded into resolved (i.e. with at least 100 star particles; $M_* \gtrsim 3.5\times10^{9}\,\Msun$)  galaxies in the simulation by means of an on-the-fly friends-of-friends galaxy finder.  If the galaxy finder finds a galaxy without a black hole, then the star particle closest to the center of mass of the galaxy is converted into a black hole, with sub-grid mass $m_\mathrm{seed} = 1.5\times10^{4}$ $\Msun$.  Note that the true particle mass remains unchanged, so that the gravitational dynamics are not impacted.  At the end of each time step, black holes within $4\rnaught$ of the potential minimum of their host dark matter halo are relocated to the center.  Here, $\rnaught$ is the smoothing length of the black hole particle.

\subsubsection{Accretion}
\label{sec:methods_black_hole_accretion}

\pkg{Simba} has a unique model for the growth of supermassive black holes (SMBH) in cosmological simulations.  It not only includes accretion via a \cite{Bondi1952} spherical accretion rate estimator, but also includes accretion that may be due to torques in cold gaseous disks near the SMBH.  The overall accretion rate is the sum of the two accretion rates

\begin{equation}
    \label{eq:overall_accretion_rate}
    \dot{M}_\mathrm{BH} = (1 - \eta)(\dot{M}_\mathrm{Torque} + \dot{M}_\mathrm{Bondi}),
\end{equation}

\noindent where $\eta = 0.1$ is the radiative efficiency of the accretion process, $\dot{M}_\mathrm{Torque}$ accretion rate due to sub-grid torques, and $\dot{M}_\mathrm{Bondi}$ is the standard Bondi accretion rate estimator.  In order to avoid double counting, the Bondi estimator is only applied to hot ($T > 10^5$ $\mathrm{K}$ gas, and the Torque estimator to cold and dense gas ($T < 10^5$ $\mathrm{K}$ and $n_\mathrm{H} > 0.13$ $\mathrm{cm}^{-3}$).  The torque estimator comes from \cite{Hopkins2011}, and models the inflow of cold gas from gravitational instabilities in sub-grid cold disks.  Additionally, the Bondi accretion rate $\dot{M}_\mathrm{Bondi}$ is limited to the Eddington accretion rate $\dot{M}_\mathrm{Edd} \equiv 4\pi GM_\mathrm{BH}m_\mathrm{p} / (\epsilon_\mathrm{r} c \sigma_\mathrm{T})$, where $M_\mathrm{BH}$ is the SMBH mass, $m_\mathrm{p}$ is the proton mass, $\epsilon_\mathrm{r} = 0.1$ is the radiative efficiency of the accretion process (see Section~\ref{sec:methods_black_hole_feedback}), $c$ is the speed of light, and $\sigma_\mathrm{T}$ is the Thomson electron scattering cross section.  The torque accretion rate $\dot{M}_\mathrm{Torque}$ is limited to $3\dot{M}_\mathrm{Edd}$ since the non-spherical nature of the accretion process could potentially lead to higher-than-Eddington rates \citep{MartinezAldama2018}.

\subsubsection{Feedback}
\label{sec:methods_black_hole_feedback}

There are two modes of feedback included in the \pkg{Simba} SMBH model: a radiative mode and a jet mode.  The radiative mode mimics highly efficient quasars with radiative efficiency $\eta \sim 0.1$ whereas the jet mode mimics the large-scale radio jets observed in massive elliptical galaxies. 

Physically, the radiative mode is expected to be dominant at accretion rates $\dot{M}_\mathrm{BH} \gtrsim 0.01 - 0.1$ $\dot{M}_\mathrm{Edd}$ \citep{Laor1989, Maccarone2003, Greene2006, McClintock2006, Sadowski2009, Madau2014a}.  In the radiative mode, gravitational energy converted in a thin accretion disk drives gas out of the local SMBH environment, leading to large scale galactic winds from the core region.  \pkg{Simba} models these winds as bipolar cold outflows and launches gas particles from the kernel region of the SMBH particle.  The velocities of these winds come from observations of X-ray detected AGN in SDSS \citep{Perna2017a} and is paramterized by the black hole mass.  At lower accretion rates ($\dot{M}_\mathrm{BH} \lesssim 0.01 - 0.1$ $\dot{M}_\mathrm{Edd}$), the environment very close to SMBHs becomes dominated by hot gas that is able to support magnetic field configurations that drive powerful, large-scale radio jets \citep{King2015}.  To model the impact of these jets, \pkg{Simba} transitions to a jet dominated regime when a SMBH reaches an accretion rate of $\dot{M}_\mathrm{BH} < 0.2$ $\dot{M}_\mathrm{Edd}$ by increasing the velocity up to $7000\,\mathrm{km}\,\mathrm{s}^{-1}$.  Additionally, SMBHs can only enter this jet dominated mode if their (sub-grid) masses reach $M_\mathrm{BH} \geqslant 10^{7.5}$ $\Msun$ in order to prevent small galaxies from being destroyed at low Eddington accretion rates.

Motivated by observations of AGN outflows and theoretical estimations, the amount of material ejected depends on the fixed momentum input $\dot{P}_\mathrm{out} = 20 L / c$ \citep{Fiore2017, Faucher-Giguere2018}.  Here, $L = \eta M_\mathrm{BH} c^2$ is the bolometric luminosity of the SMBH, $\eta = 0.1$, and $c$ is the speed of light.  Gas particles are selected stochastically to satisfy the predicted outflow rate and are decoupled from the hydrodynamics for a short period of time ($10^{-4}$ $t_\mathrm{H}$, where $t_\mathrm{H}$ is the Hubble time).

Finally, when the jet mode is active, there is an additional X-ray heating term motivated by \cite{Choi2012a}.  In galaxies with low gas fraction ($f_\mathrm{gas} < 0.2$), energy is added within the SMBH kernel following the quasar average spectra computation in \cite{Sazonov2005}.  If the gas particle to be heated is within the interstellar medium, half of the energy is added in kinetic form and the other half in thermal form to avoid extreme radiative losses.  Note that the X-ray heating only occurs when the full jet velocity is reached.

\subsubsection{Dynamics} \label{sec:methods_black_hole_dynamics}

The \pkg{Simba} model includes a mechanism for black holes to merge.  The physical assumption is that dynamical friction sufficient to keep any black holes seeded near the potential minima of galaxies during their evolution.  For that reason, black hole particles that are within $4\epsilon_\mathrm{grav,bh}$ (where $\epsilon_\mathrm{grav,bh}$ is the black hole softening size) of the potential minima of any friends-of-friends groups are repositioned to that nearest potential minimum.  The black hole velocity is reset to the velocity to the center of mass velocity of the group.  If any black hole particles are within their respective smoothing lengths $\epsilon_\mathrm{grav,bh}$, they are merged if their relative velocity is less than $3v_\mathrm{esc}$, where $v_\mathrm{esc}$ is the escape velocity.

\subsection{Galaxy and Halo Finding} \label{sec:methods_galaxy_halo_finding}

For the analysis in this work, I use the \pkg{Caesar}\footnote{\url{https://caesar.readthedocs.io/}} package,  a FoF galaxy and halo finder.  In particular, \pkg{Caesar} is able to find galaxies by searching for groups of stars and cold gas ($n_\mathrm{H} >0.13 \, \mathrm{cm}^{-3}$, $T < 10^{5}\,\mathrm{K}$) and associates them with the nearest FoF dark matter halo minimum potential center.  It is able to track both central and satellite galaxies, although it is unable to provide detailed sub-structure information or merger trees. 

\subsection{Measurements} \label{sec:methods_measurements}

There are several quantities that I use throughout this work that require definition in order to compare to the observations.  I do not attempt to apply any observational realism to the simulations throughout my analysis, instead I only focus on broad comparisons between the raw simulation data and the reported observational values.  I leave the most detailed comparison to future work.

The important quantities for the simulated galaxies are the stellar masses $\Mstar$, the effective radii $\Reff$, and the star formation rates $SFR$.

All of the simulated stellar masses that I quote in this work are defined as the FoF mass as given by the \pkg{Caesar} package (see Section~\ref{sec:methods_galaxy_halo_finding}).  For the massive galaxies that I discuss in this work, the FoF mass is within a factor of two of the stellar mass within the sphere $R \leqslant 30\,\ckpc$ centered on the star particle with the minimum potential value, at all redshifts.  I use no fixed, \textit{physical} apertures in my calculations.  Any simulated effective radius $\Reff$ of the galaxies that I quote in this work are the stellar half mass radius, i.e. the radius that encompasses 50\% of the stellar mass $\Mstar$.

There are two SFRs that I use in this work: the instantaneous ($\dot{\rho}_\mathrm{*}$; see equation~\ref{eq:h2_star_formation_rate}) and that averaged over $\Delta t = 100\,\Myr$. I use the instantaneous SFR whenever I compute the sSFRs, depletion timescales, or the impact of stellar feedback.  I use the average $SFR$ when I compute the SFR density of over the entire zoom region in Section~\ref{sec:galaxy_sfrs}.  Specifically, I save the initial mass and formation times of the star particles in my simulations and bin those values as a function of time, allowing me to compute $SFR$.  Then, I apply a filter over $\Delta t$ to smooth the rates to apply my fitting procedure, thus capturing the broad trend.

One additional quantity that I use throughout this work is the gas mass from a galaxy.  Whenever I quote a gas mass, I am referring to the cold gas mass within the galaxy, i.e. gas which is $T < 10^{5}\,\mathrm{K}$ and $n_\mathrm{H} > 0.13\,\mathrm{cm}^{-3}$.  The cold gas mass is that associated to the galaxy through the FoF procedure in \pkg{Caesar}.

\section{Accelerated galaxy evolution}
\label{sec:galaxy_populations}

\begin{figure*}
    \centering
    \includegraphics[scale=0.8]{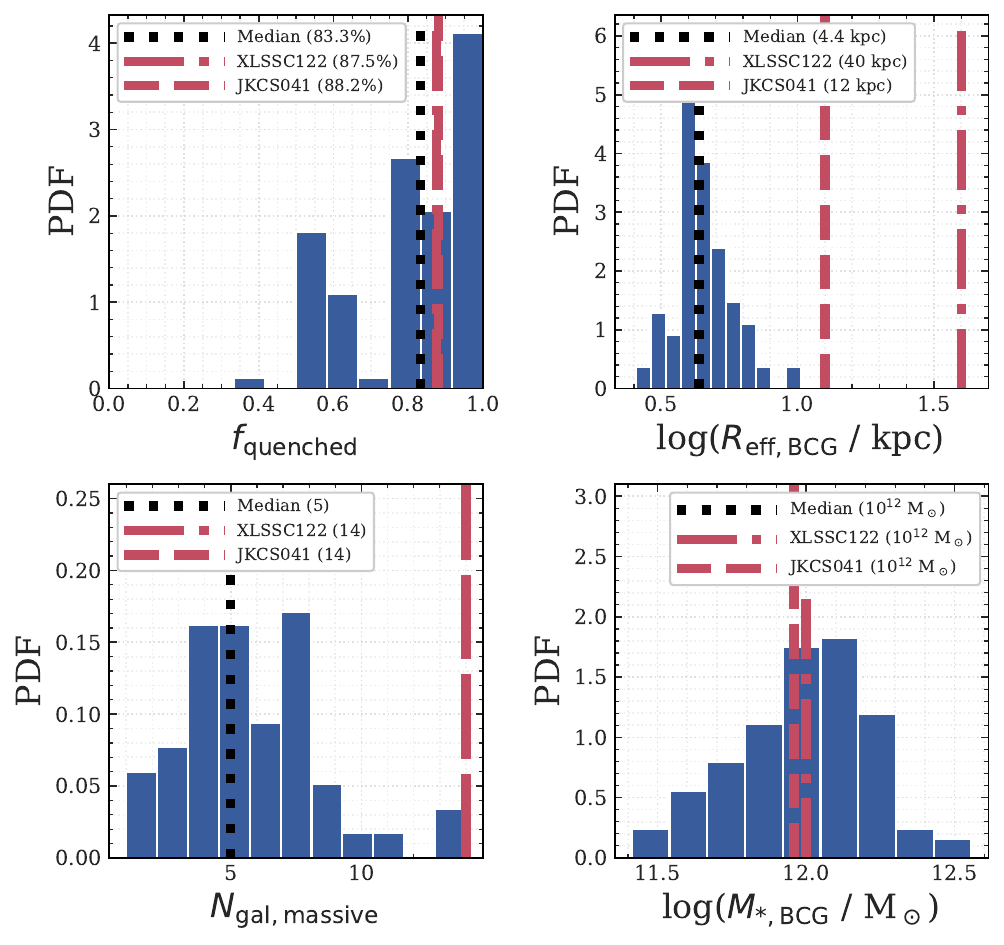}
    \caption{The histograms of brightest cluster galaxy properties at $z = 2$ from the $100$ simulated regions.  The dotted line in all plots show the simulation median, the dashed-dotted line shows XLSSC122, and the dashed line shows JKCS041. lines show the observed galaxy clusters. The quenched fraction of massive galaxies ($\Mstar>10^{10}\,\Msun$) within the core ($<0.5\rfive$) of the galaxy cluster (top left panel).  The stellar half mass radii of the brightest cluster galaxies, in physical $\kpc$ (top right panel).  The number of massive ($\Mstar > 10^{10}\,\Msun$) galaxies within the core ($<0.5\rfive$; bottom left panel). The stellar masses of the brightest cluster galaxies (bottom right panel).  The simulated galaxy population matches the observed, high redshift galaxy cluster properties. However, the brightest cluster galaxies are too compact, and there are too few massive galaxies in the cores of the simulated clusters.}
    \label{fig:quenched_reff_gal_bcg}
\end{figure*}

My sample of $100$ highly-overdense protocluster regions allows for population statistics analysis of the galaxy population down to a redshift of $z = 2$, and out to a radius of $3\,R_\mathrm{vir}$ for each re-simulated halo.  That region includes the cluster core, as well as the in-fall region where galaxy groups may be accreting.  In the following sections, I will present and discuss the BCG properties and their evolution in stellar mass, SFR, and their black hole population.  In future work, I will delve deeper into the greater galaxy population.

\subsection{Massive galaxies at z = 2}
\label{sec:galaxy_stellar_masses}

In order to test the validity of using \pkg{The Manhattan Suite} to study the evolution of high redshift overdensities, I must compare to the available observations of galaxy clusters at $z = 2$.  To that end, I use the two extreme systems JKCS041 \citep{Andreon2009, Andreon2014} and XLSSC122 \citep{Willis2020} as case studies for comparison.  

JKCS041 is spectroscopically confirmed at $z = 1.803$ and was the first high-redshift cluster to be identified above $\masstwo > 10^{14}$ $\Msun$ using $4$ different mass indicators.  In particular, JKCS041 roughly has a mass of $\masstwo \sim 2\times10^{14}$ $\Msun$ which is equivalent to my most massive halo at $z = 2$ in the simulation suite.  The observed cluster contains an upper limit of 14 red sequence, massive ($\Mstar \gtrsim 10^{10}\,\Msun$) quenched galaxies within its core ($\sim 300\,\kpc$), while the cluster itself extends out to an estimated $\rtwo \sim 700\,\kpc$.  At the core of JKCS041 there is an extremely massive ($\Mstar \sim 10^{12}\,\Msun$) BCG with an effective radius\footnote{Note that the observed $\Reff$ is based on the integrated flux, i.e. it is the radius that contains half of the integrated flux along the semi-major axis.} $\Reff \sim 10\,\kpc$.

XLSSC122 is currently the highest redshift, spectroscopically confirmed galaxy cluster, at $z = 1.98$.  The estimated mass of the cluster is $\masstwo \sim 10^{14}\,\Msun$, assuming that $\masstwo \sim 1.4\massfive$ where $\massfive = 6.3\times10^{13}\,\Msun$ is observed\footnote{The factor of $1.4$ assumes uniform density within both $\rtwo$ and $\rfive$.}.  Similarly to JKCS041, there are also $14$ massive, quenched galaxies within the core region.  The central galaxy in XLSSC122 is also $\Mstar \sim 10^{12}\,\Msun$ but has a much larger effective radius than the object in JKCS041 at $\Reff \sim 40\,\kpc$.

In both clusters, the stellar effective radii were computed by using S\'{e}rsic profile fitting to determine the half-light radius.  It is important to consider this difference since \cite{deGraaff2022} show that the half-light radius could be much higher than simulated half-mass radii.  The stellar masses of the JKCS041 and the XLSSC122 BCGs were determined using stellar population synthesis modelling, but by using different filters ($H$-band for JKCS041; F$105$W \& F$140$W for XLSSC122; see Methods of \citealt{Willis2020}).

Recall that I selected my galaxy clusters to have a halo mass of $\Mvir > 10^{14}\,\Msun$ at $z = 2$.  Therefore, both JKCS041 and XLSSC122 exist within the range of my suite.  To compare the observed clusters to the simulated clusters, I compute the BCG stellar masses and effective radii using \pkg{Caesar}. Additionally, I compute both the number of massive galaxies within the core and the quenched fraction of those galaxies.  As a first step, I compare these four quantities to the observed systems at $z = 2$.

Fig.~\ref{fig:quenched_reff_gal_bcg} shows histograms of cluster properties at $z = 2$.  The top-left plot shows a histogram of the quenched fraction ($\mathrm{sSFR} < 10^{-10.5}\,\mathrm{yr}^{-1}$) of massive ($\Mstar > 10^{10}\,\Msun$) galaxies within $0.5\,\rfive$ of all $100$ galaxy clusters.  The top-right plot shows the effective radii histogram of the BCGs, where $\Reff$ is the stellar half mass radius.  The bottom-left plot shows histogram of the number of massive ($\Mstar > 10^{10}\,\Msun$) galaxies within $0.5\,\rfive$.  Finally, the bottom-right plot shows the histogram of BCG stellar masses at $z = 2$ in all $100$ galaxy clusters.  In each plot, I overlay the median values from the simulated dataset, XLSSC122, and JKCS041 as a dotted line, dashed line, and dashed-dotted line, respectively.

First, the median quenched fraction (top left) of the simulated galaxies within the core of the clusters matches the median quenched fractions from both observed clusters.  The distribution is obviously non-Gaussian, but is skewed toward higher quenched fractions in the core regions.  There are some simulated clusters that have only half of their galaxies quenched in the core.  The galaxies in the core regions typically live in very dense ($n_\mathrm{H} \sim 0.03\,\mathrm{cm}^{-3}$) environments, and the gas temperatures in the simulated core regions are typically $T \sim 10^8\,\mathrm{K}$.  Given the extreme environment, environmental quenching acts quickly to quench these galaxies.

The bottom-left plot shows the distribution of massive galaxies within the core regions of the simulated galaxy clusters.  Here, the simulated clusters diverge from the observations significantly.  Most of the simulated galaxy clusters do not have $\mathrm{N}_\mathrm{gal,massive} > 10$ as the median is $\mathrm{N}_\mathrm{gal,massive} \sim 5$.  There are only a few clusters that have double-digit massive galaxies in the core.  That suggests one of two things: (i) the simulated galaxies in the core region are under-massive due to over-quenching or (ii) the observed galaxy clusters consist of a rare set within the rare set.  Specifically, the observed clusters could form the tip of the already rare distribution.  However, the median number of total galaxies within $0.5\rfive$ of the simulated clusters is $N_\mathrm{total,median} = 29.5$.  On average, only about $10$ of these galaxies need to grow above $\Mstar = 10^{10}\,\Msun$ for the simulations to match the observations. 

The top-right plot shows the histogram of effective radii (stellar half mass radii) of the simulated BCGs at $z = 2$.  As in the $N_\mathrm{gal,massive}$ results, the effective radii of the BCGs are under the observed effective radii by up to an order-of-magnitude.  The median value of the simulated regions is $\sim 4.4\,\kpc$ which is very compact for such massive galaxies.  While JKCS041 is closer to the median (within a factor of $\sim3$, the XLSSC122 system has an enormous $40\,\kpc$ effective radius that is an order-of-magnitude larger than the simulation set median.  This is a key tension between the simulated BCGs and the observed systems.

The bottom-right plot shows the histogram of BCG masses across all of the $100$ simulated clusters.  The median of the simulated BCGs and the observed BCGs all align at approximately $10^{12}\,\Msun$.  The stellar masses of the simulated BCGs extend out to a maximum of $3\times10^{12}\,\Msun$ and a minimum of $3\times10^{11}\,\Msun$. 

Overall, the observed BCGs in XLSSC122 and JKCS041 are consistent with the results in \pkg{The Manhattan Suite}, except for (a) the effective radii and (b) the number of massive galaxies within $0.5\,\rfive$.  That suggests that star formation may be too centrally concentrated in the BCGs and, simultaneously, there may not be sufficient star formation in satellite galaxies.  I posit that a rebalance of star formation between the central galaxy and the satellites could lead to an increase in the effective radius of the BCG, since some of the massive galaxies in the core would accrete onto the BCG -- increasing the effective radius.  Additionally, overly-concentrated galaxies are a known problem in simulated massive galaxies (see e.g., \citealt{RyanJoung2009}) and usually indicate that localized feedback in the interstellar medium is insufficient to support the gas against instability \citep{Dubois2016, Choi2018, Parsotan2020}.

\begin{figure}
    \centering
    \includegraphics[scale=0.7]{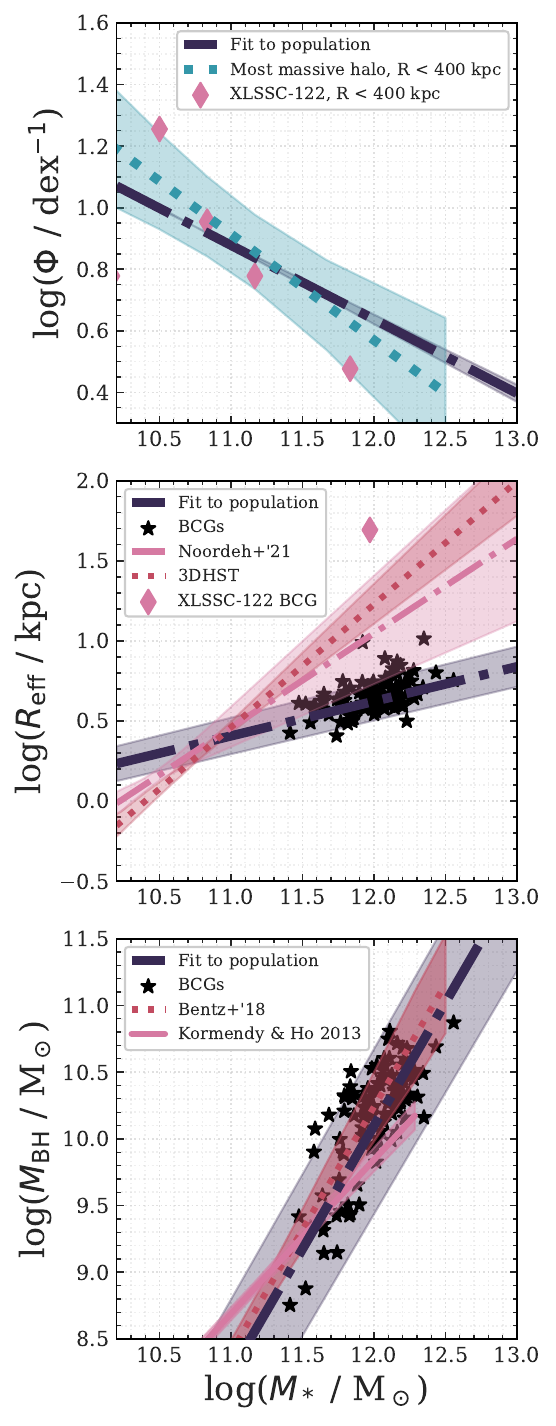}
    \caption{The typical galaxy stellar mass function within $<400\,\kpc$ of the simulated galaxy clusters (top panel), normalized per cluster, at $z = 2$.  The diamond markers show the observed XLSSC122 mass function.  The dashed-dotted line shows the typical power law fit to the mass function across all $100$ simulated galaxy clusters.  The dotted line shows the mass function within the most massive simulated galaxy cluster at $z = 2$.  The star markers show the individual $100$ brightest cluster galaxies at $z = 2$, and their stellar half mass radii as a function of their stellar masses (middle panel).  The dotted and dashed dotted lines show the observed relationships from the field and XLSSC122.  The simulated galaxies are too compact, relative to the observed cluster.  The central black hole masses of the simulated brightest cluster galaxies as a function of their stellar masses (bottom panel).  The dotted line shows two local observed relationships.}
    \label{fig:GSMF_Msigma_Reff_panel}
\end{figure}

The joint-distributions of BCG properties also provide insight into the simulated population compared to the observed population.  Fig.~\ref{fig:GSMF_Msigma_Reff_panel} shows the galaxy stellar mass function (SMF) (top), effective radii (middle), and central black hole masses (bottom) as a function of the BCG stellar masses.  In all three plots, I fit the function

\begin{equation}
    \label{eq:power_law_fit}
    \log_\mathrm{10}(y) = a\,\log_\mathrm{10}(x)+b,
\end{equation}

\noindent where $y$ is the abscissa, $x$ is the ordinate, $a$ is the slope and $b$ is the normalization, to each cluster galaxy population $i$ at $z = 2$.  I then find the distribution of parameters $(a_\mathrm{i}, b_\mathrm{i})$ and fit a Gaussian distribution to each, with means and standard deviations $\{(\mu_\mathrm{a}, \mu_\mathrm{b}), (\sigma_\mathrm{a}, \sigma_\mathrm{b})\}$, respectively.  I show the values from this procedure in Table~\ref{tab:power_law_parameters}, in Appendix~\ref{app:power_law_fits}.  I use a $95\%$ confidence interval as the error on the mean, and show that as a shaded region in each panel.

The top plot in Fig.~\ref{fig:GSMF_Msigma_Reff_panel} shows the cluster SMF.  I compare the SMF per cluster since normalizing by the volume introduces large error as the correspondence between observed volume and simulated volumes is uncertain.  I computed the SMF within $400\,\kpc$ which is effectively $\rtwo$ of XLSSC122.  The dashed-dotted line shows a power law fit using the median parameters estimated from each individual simulated galaxy cluster, as well as the error on the fit.  The dotted line shows the power law fit in the most massive simulated cluster, and error on the fit.

The diamond markers show the XLSSC122 data that I bin and compute.  It is important to compare the slopes of each distribution.  The XLSSC122 data fits well within the error of the most massive halo, but deviates significantly from the overall population.  XLSSC122 appears to have more galaxies with $\Mstar < 10^{11}\,\Msun$ and fewer galaxies with $\Mstar > 10^{11}\,\Msun$, compared to the simulated clusters.  The difference between the cluster SMFs is important since, in both cases, there has only been $4\,\Gyr$ since the Big Bang to grow and quench these massive galaxies.  The timing difference between the stellar growth and quenching time scales will be investigated in future work, but is an important tension to understand our theoretical models.

The middle plot in Fig.~\ref{fig:GSMF_Msigma_Reff_panel} shows the sizes of the BCGs as a function of their stellar masses.  The dotted line shows the size-mass relationship from a field sample (3DHST;  \citealt{Momcheva2016}), the dashed-dotted salmon line shows size-mass relationship within XLSSC122 \citep{Noordeh2021}, the dashed-dotted blue line shows the mean parameters of the fits to the galaxy population in each simulated cluster, and the black star points are the BCGs from each of the simulated clusters.  The diamond marker shows the extreme size ($40\,\kpc$) of the XLSSC122 BCG.  All of the fits were done on galaxies with $\Mstar > 2\times10^{10}\,\Msun$.  \cite{Noordeh2021} notes that the XLSSC122 sizes are not significantly different than the field, which is concerning given that the simulated massive galaxy population within my clusters are a factor of $2$ to $3$ times too compact. 

The bottom plot in Fig.~\ref{fig:GSMF_Msigma_Reff_panel} shows the central SMBH mass as a function of stellar mass in each of the $100$ simulated BCGs at $z = 2$.  The dashed line is the fit (with error) from \cite{Bentz2018} that measured local SMBH masses using reverberation mapping, the solid salmon line is the fit (with error) in \cite{Kormendy2013}, and the blue dashed-dotted line is the median fit taken from all fits to the entire cluster galaxy population within all $100$ simulated clusters.  The black stars show the individual $100$ BCGs.  Interestingly, there is no significant difference between the local observed relationships and the simulated relationship and, importantly, the BCGs also are within the error of the simulated population and the results in \cite{Bentz2018}.

The reason for this trend is that is, in fact, how the \pkg{Simba} operates.  Fig. 13 in \cite{Dave2019} shows that the \pkg{Simba} model should have SMBHs that are growing rapidly due to sustained accretion in the torque accretion model due to high gas densities.  The galaxy clusters in \pkg{The Manhattan Suite} are still growing very rapidly in the early Universe, and overshoot the \cite{Kormendy2013} relationship as they continue in an early-growth phase.  Note that Fig. 13 of \cite{Dave2019} shows that some of the quenched galaxies lay above the relationship and that my BCGs in the Manhattan are just beginning to quench after $z \sim 4$.  The SMBH masses that \pkg{Simba} predicts in this regime are much higher than the most massive in \cite{Bentz2018}, and reach the upper limit of masses in the \cite{Kormendy2013} sample.  However, the number densities of the halos in the \pkg{Manhattan Suite} are $n \sim 6\times10^{-8}\,\cMpc^{-3}$ and, therefore, one would need a complete sample of accurately measured SMBH masses out to $z \sim 0.035$ to have a single one of these halos represented\footnote{Using my assumed cosmology.}.  Indeed, they represent a rare population that should have atypical SMBH masses and, perhaps, the most massive in the Universe.  Despite the fact that they are rare, the SMBH masses predicted in \pkg{Simba} could be a factor of a few too high since I am pushing the model to its limits.

\subsection{Star formation}
\label{sec:galaxy_sfrs}

\begin{figure}
    \centering
    \includegraphics[scale=0.8]{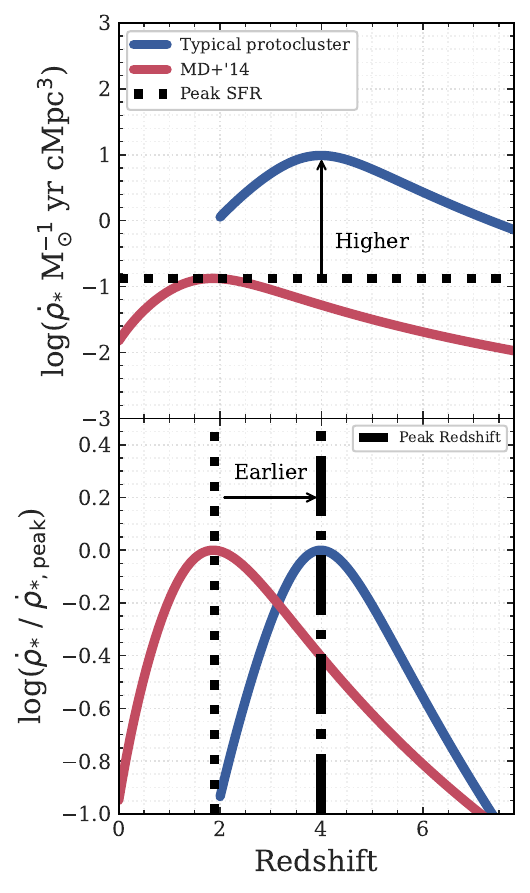}
    \caption{The star formation rate density ($\dot{\rho}_*$) of the simulated protoclusters compared to the observed field.  The higher solid curve shows the evolution of $\dot{\rho}_*$ for typical protocluster region (top panel).  The dotted line shows the peak value of the observed fit from \citealt{Madau2014}.  The simulated protocluster regions have a roughly $100\times$ peak SFR value.  The same as the top, except that each curve has been normalized to its peak value (bottom panel).  The dotted line shows the peak redshift of the field, and the dashed-dotted line shows the peak redshift of the typical protocluster region.  $\dot{\rho}_*$ peaks $\sim3\,\Gyr$ early in protocluster regions compared to the field.}
    \label{fig:sfrd_typical}
\end{figure}

The central galaxies in my simulated clusters are extremely massive at $z = 2$, with a typical mass $\Mstar \sim 10^{12}\,\Msun$ -- matching the observed BCGs in XLSSC122 and JKCS041.  If we were to assume that these galaxy grew at a constant rate since the moment after inflation, the galaxy would have to have a sustained SFR of approximately $250\,\Msunyr$ over $4\,\Gyr$!  Obviously, not all of the stellar mass grows in-situ as there should be some ex-situ contribution through mergers over cosmic time as the cluster assembles.  Not all of the galaxies in the cluster will reach the BCG, some will fall in on stable orbits and remain as satellites moving through the ever-growing halo.  The overall SFR in the protoclusters is an interesting diagnostic into the broader galaxy population, and how it evolves over time.

To examine the broader protocluster population, I compute the total SFR density of each protocluster region from $z = 8$ to $z = 2$.  The definition of protocluster region is important -- I define the protocluster region to be the comoving region that encompasses the virial radius of the $z = 2$ galaxy cluster.  For my galaxy cluster sample, a typical virial radius is $\Rvir \sim 1\,\cMpc$, so a protocluster region is typically $2\,\cMpc\times 2\,\cMpc$ on-sky, or having a volume $V_\mathrm{comoving} \sim 75\,\cMpc^3$.  Note that my definition differs subtly from the common definition where the protocluster region extends out to include anything that falls into the galaxy cluster at $z = 0$.  However, I find that definition highly biased toward $z = 0$ since the objects in my sample are already galaxy clusters based on their virial masses.  It is also important to note that the most massive galaxy clusters at $z = 2$ will not necessarily become the most massive at $z = 0$ \citep{Remus2023}.

Fig.~\ref{fig:sfrd_typical} shows the SFR density (SFRD) of a typical protocluster region compared to the observed field rate from \cite{Madau2014}.  The top plot shows the SFRD, $\dot{\rho}_\mathrm{*}$, as a function of redshift for the typical protocluster (higher) compared to the observed field (lower).  The dotted curve shows the peak value for the field. The bottom plot shows the same curves, except that I scale the SFRD to the respective peak values.  The dotted line shows the peak redshift for the field, and the dashed-dotted line shows the peak for the simulated protocluster regions.

\begin{figure}
    \centering
    \includegraphics[scale=0.7]{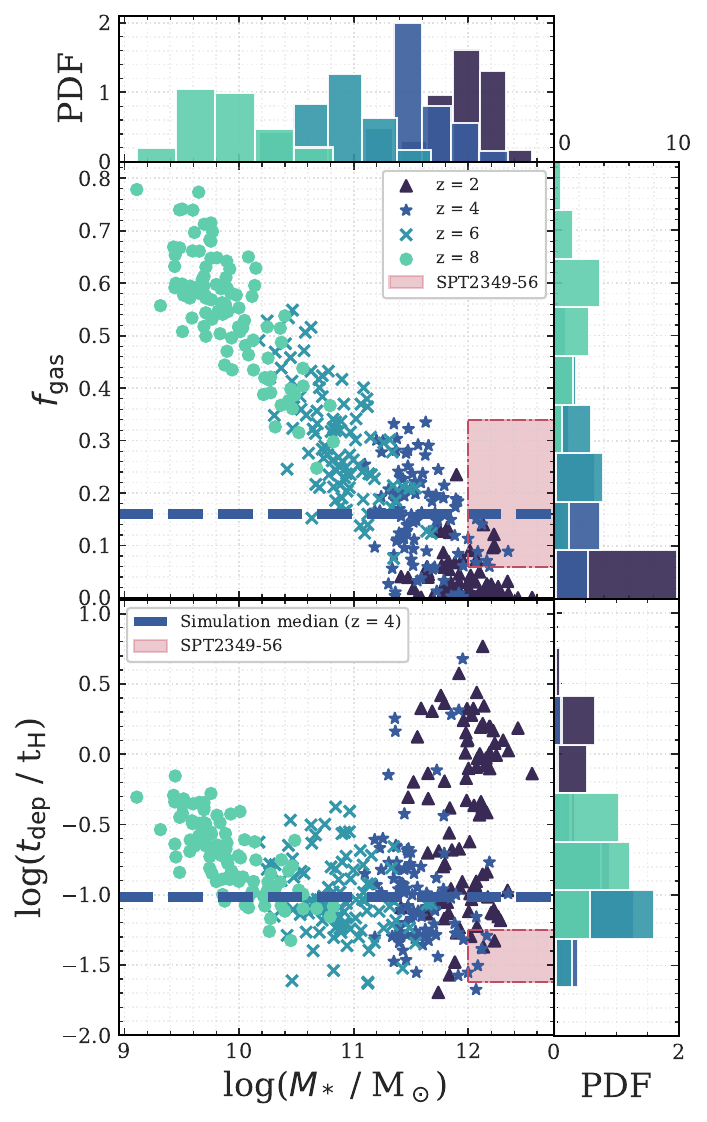}
    \caption{The evolution of the gas fractions (top) and depletion times (bottom) in the most massive progenitor branch of the individual $100$ simulated brightest cluster galaxies at $z = 8, 6, 4, 2$ -- circles, `x', stars, and triangles, respectively, as a function of stellar mass. The dashed blue line shows the simulated median of each quantity at $z = 4$, and the dashed-dotted red line shows the SPT2349-56 median observations from \citealt{Hill2022}.  The very top plot shows the histograms of stellar mass at the same redshifts.  The farthest right plots show the histograms of gas fraction (top right) and depletion times (bottom right).  The gas fractions and depletion times match the SPT2349-56 object, and suggest that the extreme observed protocluster is an outlier of the already rare objects.}
    \label{fig:galaxy_evolution_fgas_tdep}
\end{figure}

To compute a single curve that represents all of my protocluster regions, I first fit the SFRD evolution of each protocluster to the cored power law profile from \cite{Madau2014},

\begin{equation}
    \label{eq:madau_dickinson}
    \dot{\rho_\mathrm{*}}(z) = a\frac{(1 + z)^b}{1 + [(1+z)/c]^d}\,\Msun\,\yr^{-1}\,\cMpc^{-3},
\end{equation}

\noindent where $z$ is the redshift, $a$ is the normalization, $b$ is the inner slope, $d$ is the outer slope, and $c$ is the core radius (in redshift).  After fitting $100$ curves, I use the mean $a$, $b$, $c$, and $d$ values as the typical protocluster region.  The best fit values from both \cite{Madau2014} and my suite are shown in Table~\ref{tab:sfrd_parameters}, in Appendix~\ref{app:power_law_fits}.  The most important differences are in the $b_\mathrm{sim} = 5.5$ and $d_\mathrm{sim} = 11.5$ slopes -- \pkg{The Manhattan Suite} shows that these rare protoclusters typically have steeper rises and falls in the SFRD density compared to the field ($b_\mathrm{MD14} = 2.7$ and $d_\mathrm{MD14} = 5.6$).  Additionally, the value $c_\mathrm{sim} = 5.0$ is higher than $c_\mathrm{MD14} = 2.9$, indicating that the peak star formation is happening much earlier.  Specifically, the peak star formation occurs approximately $1.7\,\Gyr$ before the peak in the field, using my assumed cosmology.

The earlier and more intense star formation in protocluster regions compared to the field relates to the fact that I select the overdensities for resimulation to be biased high.  I argue that this is exactly what happens in the observations, as there is a bias to selecting the brightest objects.  It is known that protoclusters have higher than average SFRDs \citep{Chiang2017}, but some of the intense starbursts have failed to be reproduced in our theoretical models of galaxy evolution \citep{Remus2023}.

\subsection{Gas fractions and depletion times} \label{sec:galaxy_gas_fractions}

One particular interesting, extreme object that is difficult to understand given our theoretical models is the protocluster region SPT2349-56 at z = 4.3.  The core of the protocluster contains $14$ star forming ($\mathrm{SFR}\sim5000\,\Msunyr$), massive galaxies ($\Mstar \sim 10^{10}\,\Msun$), all within a $130\,\kpc$ physical region \citep{Hill2022}.  There have been other cosmological simulations that attempt to recover the specific SPT2349-56 object, but were unable to match the depletion times, $t_\mathrm{dep} = \Mgas / \mathrm{SFR}$ and, hence, the gas fractions, $f_\mathrm{gas} = \Mgas / (\Mgas + \Mstar)$, within the system \citep{Yajima2021, Remus2023}.  Nor has the richness been possible to reproduce.  Obviously, SPT2349-56 is a special protocluster (i.e. special among extreme protoclusters) but the goal of my project is to find rare-of-rare objects in rare overdensities at high redshift.

In \cite{Rennehan2020}, we argued that the $14$ originally-observed SPT2349-56 galaxies would quickly form a BCG (within $t \sim 300\,\Myr$).  I posit that these BCGs are the same BCGs I discussed in Section~\ref{sec:galaxy_stellar_masses}, that are in the observed systems XLSSC122 and JKCS041.  To examine this possibility, I must examine the depletion times and gas fractions and compare to the SPT2349-56 system.

Fig.~\ref{fig:galaxy_evolution_fgas_tdep} shows the gas fractions (top) and depletion times (bottom) for the most massive progenitor branches of the $100$ simulated BCGs, at redshifts $z = 2, 4, 6, 8$, given by the triangles, stars, crosses, and circles, respectively.  The histograms at the top of the figure shows the stellar mass histograms of the BCGs from $z = 2, 4, 6, 8$ from darkest to lightest, respectively.  Similarly, the two histograms on the right of the figure show the $f_\mathrm{gas}$ (top right) and $t_\mathrm{dep} / \mathrm{t}_\mathrm{H}$ histograms from snapshots at $z = 2, 4, 6, 8$, from darkest to lightest, respectively.  Here, $t_\mathrm{H}$ is the Hubble time \textit{at that particular redshift}.  In both central plots, there is a dashed line which represents the simulation median at $z = 4$ of either $f_\mathrm{gas}$ (top) or $t_\mathrm{dep} / \mathrm{t}_\mathrm{H}$ (bottom).  The shaded regions show the upper and lower bounds for $f_\mathrm{gas}$, $t_\mathrm{dep}$, and $\Mstar$ in SPT2349-56 at z = 4.3 \citep{Hill2022}. 

First, it is obvious that the gas fractions (top plot) of the most massive progenitors of the BCGs are constantly decreasing from $z = 8$ to $z = 2$.  By $z = 2$, the BCGs have effectively zero gas within their cores, at least compared to their stellar masses.  At $z = 4$, the median gas fraction in the BCGs is $\overline{f}_\mathrm{gas,sim} \sim 0.16\pm 0.08$ whereas the SPT2349-56 galaxies have a median of $\overline{f}_\mathrm{gas,obs} \sim 0.20\pm0.14$.  The error on each median represents a single standard deviation of the distribution, not including observational error.  The medians are consistent, and surprisingly close given that I select galaxy clusters at $z = 2$ without knowing whether they reproduce the extreme protocluster regions observed in the early Universe.  

The bottom plot in Fig.~\ref{fig:galaxy_evolution_fgas_tdep} shows the depletion time scaled to the Hubble time at each given redshift $z = 2, 4, 6, 8$ -- darkest to lightest, respectively, for all of the $100$ simulated BCG progenitors.  At $z = 8$ (circles), there is already a wide spread in depletion times and all of the most massive progenitors have depletion times that are less than a Hubble time ($t_\mathrm{H}|_{\mathrm{z=8}} \approx 0.64\,\Gyr$) with a mean value $t_\mathrm{dep} / \mathrm{t}_\mathrm{H} = 0.23\pm0.13$.  At $z = 6$, the mean depletion time is $t_\mathrm{dep} / \mathrm{t}_\mathrm{H} = 0.12\pm0.08$ and, at $z = 4$, the mean depletion time is $t_\mathrm{dep} / \mathrm{t}_\mathrm{H} = 0.23\pm0.58$.  At $z = 2$, the mean depletion time rises sharply to $t_\mathrm{dep} / \mathrm{t}_\mathrm{H} = 0.78\pm0.95$, and many galaxies move to values longer than a Hubble time. The mean depletion time of the SPT2349-56 galaxies is much lower than the simulated mean values, at $t_\mathrm{dep} / \mathrm{t}_\mathrm{H} = 0.04\pm0.016$.  While the mean values are not close, the wide spread in values of the simulated distributions allows the two values to have overlap within the $1\sigma$ spread.  The SPT2349-56 system seems to represent the edge cases of the $100$ most massive progenitors of the BCGs in \pkg{The Manhattan Suite}.  In particular, there are only $4$ proto-BCGs that are in \textit{both} of the shaded regions for SPT2349-56 of Fig.~\ref{fig:galaxy_evolution_fgas_tdep}.

\subsection{Quenching}
\label{sec:galaxy_quenching}

The simultaneous decrease in gas fractions and increase in depletion timescales of the progenitor brightest cluster galaxies (BCGs) at $z = 4$ indicates that some process is impacting the build-up of gas into the core region of the protoclusters.  That process may be consumption, i.e. all of the gas is converting into stars, stellar feedback, AGN feedback, gravitational heating, or some combination of these four processes.  It is unlikely that stellar feedback is the culprit given the low energy injection rate into the surrounding medium, and the depth of the potential in these massive halos. 

First, let us investigate the SFRs in the progenitor galaxies at the same redshifts as in Section~\ref{sec:galaxy_sfrs}, $z = 2, 4, 6, 8$.  

\begin{figure}
    \centering
    \includegraphics[scale=0.7]{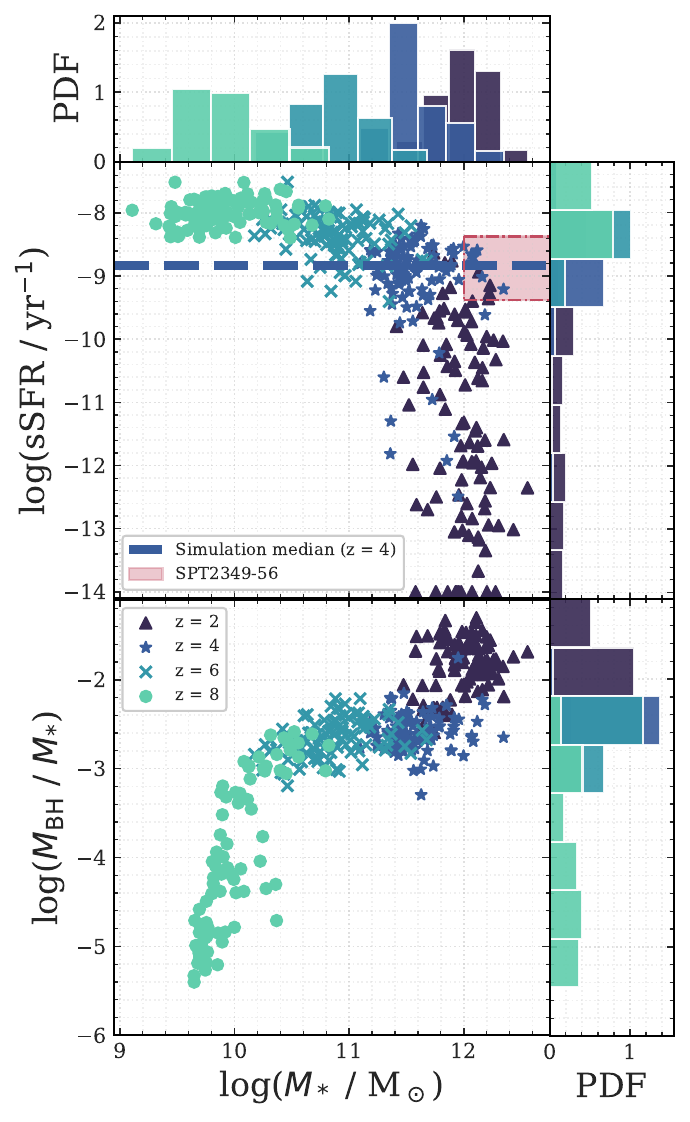}
    \caption{The evolution of the specific star formation rate (top) and central black hole masses (bottom) in the most massive progenitor branch of the individual $100$ simulated brightest cluster galaxies at $z = 8, 6, 4, 2$ -- circles, `x', stars, and triangles, respectively, as a function of stellar mass.  The dashed blue line in the top plot shows the median specific star formation rate of the $100$ most massive progenitors at $z = 4$, and the shaded region shows the SPT2349-56 estimated specific star formation rate in \citealt{Hill2022}.  Most of the simulated galaxies quench after $z = 4$, and match the observed specific star formation rate in SPT2349-56.  The very top plot shows the histograms of stellar mass at the same redshifts.  The farthest right plots show the histograms of the specific star formation rate (top right) and central black hole masses (bottom right).}
    \label{fig:galaxy_ssfr_bh_evolution}
\end{figure}

Fig.~\ref{fig:galaxy_ssfr_bh_evolution} shows the evolution of the sSFR ($\mathrm{sSFR}$; top plot) and the central SMBH mass-stellar mass ratio (bottom plot) as a function of stellar mass, for the redshifts $z = 2, 4, 6, 8$ -- from darkest to lightest, respectively.  The top histogram shows the stellar masses distributions for the same redshifts, whereas the two histograms on the right side of the figure show the $\mathrm{sSFR}$ (top right) and $M_\mathrm{BH} / M_\mathrm{*}$ (bottom right) histograms.  In the $\mathrm{sSFR}$ panel, I include a comparison between the simulation median at $z = 4$ (dashed line) to the SPT2349-56 estimated $\mathrm{sSFR}$ in the banded region.  Note that the shaded region for SPT2349-56 extends from assuming either $10^{12}\,\Msun$ or $10^{13}\,\Msun$ for the total enclosed stellar mass in the system, as per \cite{Hill2022}. 
 Immediately striking is the transition in both quantities from $z = 4$ to $z = 2$ that mimics the sudden increase (over the same time period) in the spread of depletion times in the bottom plot in Fig.~\ref{fig:galaxy_evolution_fgas_tdep}.

First, let us examine the $\mathrm{sSFR}$ evolution.  Evidently, the BCG progenitors form stars rapidly at $z = 8$ and the star formation gradually decreases by an order of magnitude at $z = 4$.  The $\mathrm{sSFR}$ distribution at $z = 2$ is much broader than higher redshifts.  Assume that galaxies quench below $\mathrm{sSFR}_\mathrm{quenched} = 0.2 / t_\mathrm{H}(z)$ \citep{Pacifici2016, Lammers2023}, where $t_\mathrm{H}(z)$ is the Hubble time at redshift $z$.  At $z = 4$, there are a few BCG progenitors that have already quenched and, at the time, their masses are typically $\Mstar \sim 3\times10^{11}\,\Msun$.  By $z = 2$, the majority of the BCGs are quenched -- a fact that is consistent with their gas fractions approaching zero (recall Fig.~\ref{fig:galaxy_evolution_fgas_tdep}).  

Second, consider the growth of $\Mbhmstar$ as a function of $M_\mathrm{*}$ in the bottom plot in Fig.~\ref{fig:galaxy_ssfr_bh_evolution}.  At $z = 8$ there is a large spread in $\Mbhmstar$, spanning two orders of magnitude from $\Mbhmstar \sim 10^{-5}$ to $10^{-3}$ with a minimal spread in stellar masses.  At $z = 8$ there is already a trend that the more massive SMBHs reside in the most massive galaxies.  From $z = 6$ to $z = 4$ there is minimal evolution in the mean values $\log_\mathrm{10} \overline{M}_\mathrm{BH} = 8.21$ to $\log_\mathrm{10} \overline{M}_\mathrm{BH} = 8.94$, which indicates that the SMBHs are growing in lockstep with their host proto-BCGs.  From $z = 4$ to $z = 2$, $\Mbhmstar$ grows almost by an order of magnitude from $\log_\mathrm{10} \overline{\Mbhmstar} = -2.63$ to $\log_\mathrm{10} \overline{\Mbhmstar} = -1.84$.  In the same time frame, the stellar masses only grow marginally from a mean value of $\log_\mathrm{10} \overline{\Mstar} = 11.6$ to $\log_\mathrm{10} \overline{\Mstar} = 12.0$.  Obviously, the SMBHs continue to grow while their host galaxies quench, which is consistent with the $\mathrm{sSFR}$ plot in Fig.~\ref{fig:galaxy_ssfr_bh_evolution}.  If the SMBHs are the cause of the reduction in gas fractions and decrease in SFRs, how do they continue to grow past the point when the galaxy gas fractions drop to zero?

It is useful to discuss the relationship between the SMBH growth rate and galaxy quenching in the context of the \pkg{Simba} sub-grid models that I outline in Section~\ref{sec:methods}.  Recall that the SMBH growth rate in \pkg{Simba} has three modes: (a) a cold accretion mode, (b) a hot, Bondi-like accretion mode, and (c) merger-driven growth.  In (a), the idea is that gravitational torques within stellar and gaseous disks on sub-grid scales drive gas toward the central region of the galaxy and into the SMBH's sphere of influence.  The key to the cold accretion mode is that it requires cold disks in order for the torques to operate.  Conversely, in (b), the Bondi accretion rate depends only on having a hot, spherically-distributed halo that collapses in a spherically symmetric manner onto the SMBH.  In \pkg{Simba}, (a) is only computed for gas that is cold ($n_\mathrm{H} > 0.13\,\mathrm{cm}^{-3}$ and $T < 10^{5}\,\mathrm{K}$) and (b) is only computed for hot gas ($n_\mathrm{H} < 0.13\,\mathrm{cm}^{-3}$ and $T > 10^{5}\,\mathrm{K}$) giving a clear demarcation for the two modes.  In cosmological simulations like \pkg{The Manhattan Suite}, cold gas (in the same definition) is responsible for star formation since the star formation directly scales with the cold gas density, i.e. $\dot{\rho}_\mathrm{*} \sim \rho_\mathrm{gas,cold} / t_\mathrm{ff}$, where $t_\mathrm{ff}$ is the local free-fall time.  Of course, all of these sub-grid models are observationally and physically motivated to represent how the Universe works on small scales.  If we assume that these models are roughly accurate, it is completely reasonable that as the cold gas fraction in the cores of the proto-BCGs decreases, the contribution to the cold accretion mode in the SMBHs also decreases in a similar manner. However, as the cold gas is either consumed, heated, or ejected, it should create a low pressure region that must be filled.  If it were to be replaced by cold gas, the galaxies would continue to form stars.  However, we know from Fig.~\ref{fig:galaxy_ssfr_bh_evolution} that the majority of the BCGs are quenched at $z = 2$.  Therefore, the gas is most likely partially consumed, ejected, and heated above $T\sim10^{5}\,\mathrm{K}$ such that (i) star formation halts and (ii) the SMBH continues to grow via Bondi accretion of hot gas.  Although I do not show the results here, channel (c) contributes little to the overall growth of the BH during the time period between $z = 4$ and $z = 2$.  Indeed, the picture matches the recent work in \cite{Sharre2024}, where they explain which astrophysical models control the SFRD history of the Universe (using the same sub-grid models as here).

\begin{figure}
    \centering
    \includegraphics[scale=1.0]{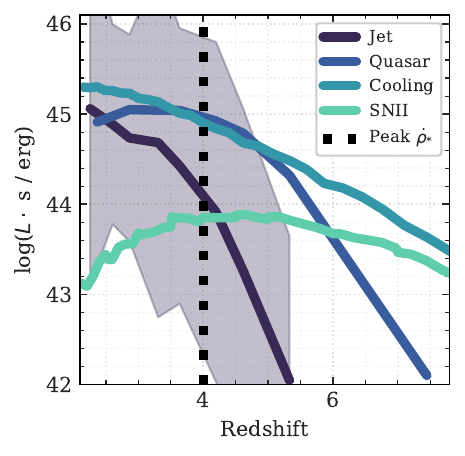}
    \caption{The average luminosity evolution across the $100$ simulated protocluster regions.  The light blue curve shows the typical mean cooling luminosity within $\rfive$, the medium blue curve shows the mean quasar luminosity of all of the central black holes in the most massive progenitors of the $z = 2$ brightest cluster galaxy progenitors, the dark blue curve shows the typical jet luminosity, and the lightest blue curve shows the contribution from type II supernova.  The shaded region shows typical $1\sigma$ jet fluctuations around the mean.  The dotted line shows the typical peak star formation rate density redshift of the protocluster regions.  The quasar luminosity and $1\sigma$ jet luminosity both cross the cooling luminosity at $z\sim5$, which leads to the decline in star formation rate density at $z\sim4$.  I predict that highly star forming protoclusters should have an active galactic nucleus with $L\sim10^{45}\,\ergs$ at $z\sim4$ in order to combat the cooling luminosity.}
    \label{fig:black_hole_energy_rate_comparison}
\end{figure}

There is nothing particularly new in this picture, as it is how SMBH feedback is expected to act to quench galaxies.  The only new component is that it happens rapidly in the high overdensities that I simulate (recall Fig.~\ref{fig:sfrd_typical}).  However, that allows me to make predictions for the types of AGN luminosities that we would expect to observe in highly-overdense protoclusters in the early Universe.  In support of both of these claims, I must compare the typical jet and quasar luminosities in the central SMBHs of the proto-BCGs to the cooling luminosity of the hot gas within $\rfive$ of the protocluster region.

Fig.~\ref{fig:black_hole_energy_rate_comparison} shows (as solid lines) the comparison between the typical jet, quasar, cooling, and SNII luminosities within the protocluster regions, from darkest to lightest, respectively.  To compute the SNII energy rate, I use the instantaneous SFR and SNII fraction from \pkg{Simba}, $f_\mathrm{SNII} = 0.18$, combined with the expected number of SNII per unit stellar mass for the \cite{Chabrier2003} IMF, $n_\mathrm{SNII} = 0.0173\,\Msun^{-1}$ (see e.g. \citealt{Vogelsberger2014}) and assume $E_\mathrm{SNII} = 10^{51}\,\mathrm{erg}$.  I show the jet and quasar luminosities only for the central SMBH in the BCGs (and their most massive progenitors).  The filled dark region surrounding the jet luminosity curve shows the $1\sigma$ region of mean jet fluctuations, as the jet is extremely stochastic on short timescales within the central region of the protoclusters.  For comparison, I show the peak SFRD from Fig.~\ref{fig:sfrd_typical} as a dotted black line at $z \sim 4$.

First, the cooling luminosity increases by roughly two orders-of-magnitude from $z = 7$ to $z = 2$ to $\dot{E}_\mathrm{cool} \sim 10^{45}\,\ergs$.  Simultaneously, the typical quasar luminosity is increasing at a comparatively accelerated rate, crossing the cooling luminosity curve at $z \sim 5$ and remaining above until $z \sim 3$.  The SNII energy rate is insufficient at all times to suppress cooling in the protocluster cores.  The jet luminosity reaches $\dot{E}_\mathrm{jet} \sim 10^{42}\,\ergs$ at $z \sim 5$ just as the quasar luminosity crosses the cooling luminosity.  The $1\sigma$ jet fluctuations also cross the cooling luminosity threshold at $z \sim 5$ but continue to increase in strength rather than the decreasing trend with the quasar luminosity.  The typical properties of the luminosities within my simulated protocluster regions supports the picture that AGN feedback is responsible for the decline of the typical SFRD after $z = 4$ since the quasar and jet luminosities conspire to cross the cooling luminosity slightly earlier than this point, at $z \sim 5$.  The quasar mode in \pkg{Simba} is intrinsically linked to the amount of cold gas within the sphere of influence of the SMBH, and the jet only becomes active for very low accretion rates when one would expect the Bondi accretion mode to become dominant.  The beginning of the downturn in the quasar luminosity occurs for the same reason the decline in SFRD occurs --- there is less cold gas in the galaxy (recall Fig.~\ref{fig:galaxy_evolution_fgas_tdep}) and the jet becomes the dominant energy injector (and continuously increases).

Given that the picture I outline is supported by the data and that the protocluster regions I simulate are good representations of extreme protoclusters like SPT2349-56, it is possible for me to predict broadly for protocluster regions at high redshift.  Specifically, the AGN feedback picture suggests that highly star forming protoclusters should, at their peak, also host AGN with luminosities somewhere in the range of $L_\mathrm{AGN} \sim 10^{45}\,\ergs$.  These AGN may be deeply embedded and shrouded by dust so may be difficult to detect.  However, \pkg{The Manhattan Suite} suggests that large-scale jet activity should be present in the protocluster population, with evidence already available at $z \sim 4$ (when they are peaking in star formation!).  Interestingly, recent observations of SPT2349-56 suggest that there is a radio-loud AGN in the most massive galaxy, with an estimated power of $L_\mathrm{AGN} \sim 10^{45}\,\ergs$ \citep{Chapman2024} --- further supporting my prediction that protoclusters should broadly host powerful AGN at $z \sim 4$.

\section{Monsters at high redshift?}
\label{sec:galaxy_monsters}

\begin{figure*}
    \centering
    \includegraphics[scale=0.8]{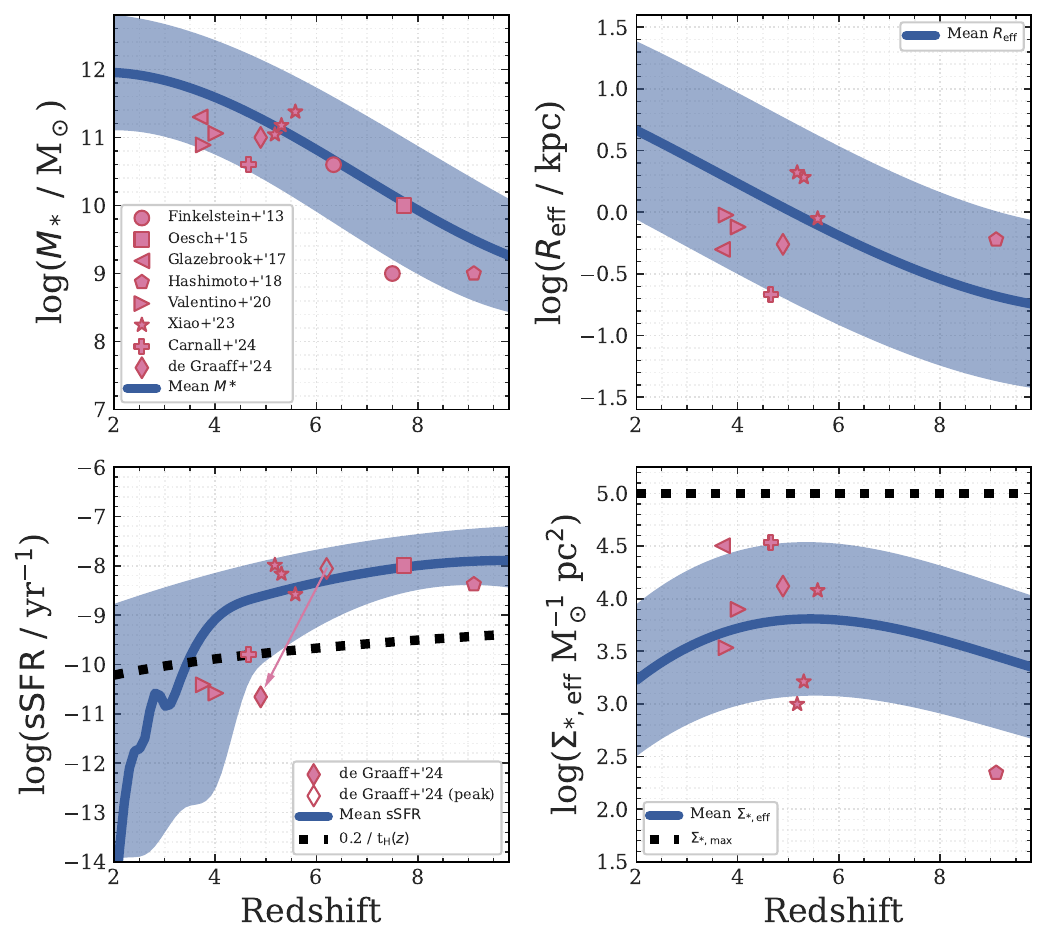}
    \caption{A comparison of the most massive progenitor branches of the $100$ simulated brighest cluster galaxies (solid blue curves) and the observed, extremely massive galaxies at high redshift (salmon markers).  The shaded blue region shows the $2\sigma$ region around the mean values for the most massive progenitor tracks.  (top left) The stellar mass evolution.  The stellar masses of the most massive progenitors act as an upper bound for the observed massive galaxies. (top right) The stellar half mass radii.  The simulated progenitors match the observed, compact nature of the high mass galaxies. (bottom left) The specific star formation rate, where the dotted line shows the boundary between quenched and star forming at $\mathrm{sSFR} = 0.2/t_\mathrm{H}(z)$.  The hollow diamond shows the estimated star burst and stellar mass of the \citealt{deGraff2024} galaxy, $\sim300\,\Myr$ pre-observation.  The observed massive galaxies match the simulated progenitors, and interestingly the estimated starburst.  (bottom right) The stellar surface density.  The dotted line shows the theoretical maximum value in \citealt{Hopkins2010b}.  I predict that the stellar surface densities should peak at $z\sim5$, just before the peak in the star formation rate density.}
    \label{fig:rare_bcgs}
\end{figure*}

In addition to extreme galaxy clusters and protoclusters at high redshift, there have been numerous detections of extreme \textit{galaxies} thanks to the power of JWST.  The fact that the BCGs in \pkg{The Manhattan Suite} reach mean masses of $\log_\mathrm{10} \overline{\Mstar} = 9.91, 10.9, 11.6, 12.0$ at $z = 8, 6, 4, 2$, respectively, suggests that the extremely massive (sometimes quenched) galaxies in the early Universe could be located in over-dense regions at high redshift.  I want to focus on the most massive progenitors of the BCGs from my sample, but want to remind the reader that there is a coeval population of massive galaxies within the core region of the protocluster.  It is important to note that I am focusing on the most extreme galaxies.  In order to make claims about broader populations of galaxies, I would need to simulate the entire host volume, i.e. $(1.5\,\cGpc)^3$, from which I sample the galaxy cluster regions at $z = 2$.  That would require approximately $(10,000)^3$ resolution elements, so is currently unfeasible\footnote{Although, see \citealt{Schaye2023} for the FLAMINGO simulations that are only an order-of-magnitude away from the necessary mass resolution in a volume that is $\sim0.3\times$ smaller than mine.}.  However, I show below that my sample of galaxy (proto-)clusters is perfectly capable of hosting the extreme observed galaxies.

Fig.~\ref{fig:rare_bcgs} shows the mean properties of the most massive progenitors of my $100$ BCGs from $z = 2$ to $z = 9$.  The top-left plot shows the stellar mass evolution, the top-right plot shows the effective radius (i.e. stellar half mass radius), the bottom-left plot shows the sSFR, and the bottom-right plot shows the stellar surface density.

First, the top-left plot in Fig.~\ref{fig:rare_bcgs} shows the mean stellar mass evolution along with the $2\sigma$ region that I compute in bins of redshift.  For comparison, I show the estimated masses and redshifts for several \textit{extreme} galaxies from, in chronological order: \cite{Finkelstein2013, Oesch2015, Glazebrook2017, Hasimoto2018, Valentino2020, Xiao2023, Carnall2024, deGraff2024}.  I do not include the observational uncertainties for clarity.  All of the observational works used spectral energy distribution modeling to compute their stellar masses and $SFR$ values.  \cite{Finkelstein2013} and \cite{Oesch2015} did not provide effective radii of their sources.  All other publications report the half-light radius based on the filter used in their work (F$160$W, \citealt{Hasimoto2018}; $K$-band, \citealt{Valentino2020}; F$277$W, \citealt{Carnall2024}; F$444$W, \citealt{Xiao2023} and \citealt{deGraff2024}), except for \cite{Glazebrook2017} who used the high-redshift mass-size relationship from \cite{Straatman2015}.  All of the extreme galaxies from the selected works agree well with the histories of the most massive progenitors of the BCGs in my sample.  

The top-right plot shows the mean effective radii of the simulated galaxies (solid line).  The markers show the same observational points as in the previous plot.  Here, I define the effective radius as the stellar half mass radius and report the results in \textit{physical} $\kpc$.  All of the observations are consistent with the $2\sigma$ spread, i.e. the shaded region around the mean curve.  

The bottom-left plot shows the mean sSFR of the progenitors of the BCGs.  Given the highly stochastic nature of the star formation, I show the minimum and maximum $\mathrm{sSFR}$ in each given time bin as the shaded region around the mean.  Additionally, I fit a cubic spline over all of the data to show the general trend against the noise.  The dotted line shows one particular estimate of quenching, where the $\mathrm{sSFR}$ must be below $\mathrm{sSFR} \lesssim 0.2 / t_\mathrm{H}(z)$ \citep{Pacifici2016}, where $t_\mathrm{H}(z)$ is the Hubble time at redshift $z$.   The markers show the same observational data as in the other plots in Fig.~\ref{fig:rare_bcgs}, except now I show the estimated starburst from \cite{deGraff2024} as a hollow diamond.  That point is $\sim 300\,\Myr$ before the observation at $z\sim5$.  All of the observations that have star formation estimates are within the minimum and maximum expected spread in $\mathrm{sSFR}$ except for the observation in \cite{deGraff2024}.  However, I do not count that as a failure of the model as the snapshot resolution obfuscates the SFRs, since a long cadence may cause me to miss strong fluctuations in either direction.  The general trend is the same for that one observation, where there is a star burst followed by quenching $\sim300\,\Myr$.  In fact, we predicted these types of galaxies in our \cite{Rennehan2020} work and match the estimated timescale in \cite{deGraff2024} accurately.

The bottom-right plot in Fig.~\ref{fig:rare_bcgs} shows the mean stellar surface density evolution as a function of redshift across the most massive progenitors of the BCGs (solid line).  The shaded region shows the $2\sigma$ variance in each time bin.  The stellar surface density is the stellar mass per unit effective radius area.  The markers show the same observations as in the rest of the plots.  The dotted line shows the estimated maximum possible stellar surface density from \cite{Hopkins2010b}.  On average, the stellar surface density evolution in the simulated proto-BCGs grows to a peak at $z \sim 5$ and declines toward $z = 2$.  Most of the observations are consistent with the spread in the simulation distributions, except that the \cite{Hasimoto2018} observation has a lower stellar surface density.

\section{Weaving the Thread}
\label{sec:discussion}

\pkg{The Manhattan Suite} is a \textit{biased} selection of $100$ galaxy clusters at $z = 2$ where the bias is intentional, on my part.  My goal is to achieve a similar selection bias as observations of high redshift galaxies, in attempt to understand how these special galaxies evolve.  In particular, \pkg{The Manhattan Suite} is \textit{biased} towards highly overdense regions of the Universe, at high redshift.  My assumption was that the density of these regions is high and, therefore, the interaction time scale is very short, leading to enhanced SFRs over the field.  Not only should SFRs be increased, but all interaction timescales should decrease in these overdense regions. Recall that the free-fall times and orbital times scale with the inverse of the density, $t_\mathrm{ff} \propto t_\mathrm{orbit} \propto 1 / \sqrt{G\rho}$, where $G$ is Newton's gravitational constant and $\rho$ is the matter density.  My selected sample has much higher densities compared to similar mass objects at $z = 0$, simply because they formed earlier in the Universe when the average density was higher.

My hypothesis is that systems such as the extremely starbursting SPT2349-56 protocluster, the high-redshift galaxy clusters XLSSC122 and JKCS041, as well as the extremely massive high-redshift galaxies are intimately linked.  They should be, in my opinion, linked through the fact that they all are undergoing \textit{accelerated} galaxy evolution because of their environment --- a combination of the halo peak and assembly biases. 

Observations of massive galaxies at high redshift usually seem inconsistent with theoretical models, as their properties are too extreme to be explained on average.  For example, the excellent summary in Section 8 of \cite{deGraff2024} demonstrates that contemporary models would predict number densities of $n\sim10^{-8}\,\cMpc^{-3}$ for galaxies such as the example in their work.  That is in agreement with my sample, since there are only roughly $\sim200$ galaxy clusters at $z = 2$ in my $(1.5\,\cGpc)^3$ parent volume ($n \sim 6\times10^{-8}\,\cMpc^{-3}$).  However, it is a much simpler explanation that the observed massive galaxies are proto-BCGs rather than they are in tension with our models.  If we accept that they are common, in terms of progenitors of BCGs, then some of the observed fields must be overdense regions.  If they are not overdense, usually that invokes the need for higher than average star formation efficiencies \citep{BoylanKolchin2023, Chworowsky2023b, Xiao2023}.  Additionally, there are also models that posit feedback-free star formation could lead to massive galaxies quickly with more reasonable star formation efficiencies \citep{Dekel2023}.  

The (proto-)BCGs in \pkg{The Manhattan Suite} do, in fact, act as if they are feedback free since the AGN feedback and the stellar feedback are unable to overcome the cooling luminosity at high redshifts (recall Fig.~\ref{fig:black_hole_energy_rate_comparison}).  While the mean stellar mass of the proto-BCGs is $\Mstar\sim10^{10}\,\Msun$ at $z\sim8$, the AGN feedback luminosity only crosses the cooling luminosity in the core of the halo at $z\sim5$.  During that period, the massive halo has no chance of suppressing the enormous cooling flow onto the galaxy.  Therefore, the simplest explanation for the observed massive, compact galaxies at high redshift is that they are (a) proto-BCGs and (b) forming in a feedback free fashion.

It is interesting that, if these massive galaxies at high redshift are also related to the SPT2349-56 object (and other SPT-selected protoclusters), then there should be much more extended structure in the regions surrounding the galaxies due to halo peak bias.  For example, there should be a population of massive galaxies surrounding the proto-BCGs.  Therefore, if a survey finds one particularly massive galaxy, one would expect more within a region of the virial radius of a typical galaxy cluster at $z = 0$, e.g. $\Rvir \sim 1 - 5\,\cMpc$, or $V \sim 1 - 125\,\cMpc^3$.  

Finding analogs of extremely starbursting protoclusters in cosmological simulations of the Universe has been a difficult endeavor. For our analysis in \cite{Rennehan2020}, we constructed an isolated simulation of SPT2349-56 based on the observations in \cite{Miller2018} and the using scaling relationships from \cite{Behroozi2013a}.  We were able to reproduce SFRs of $\sim3000\,\Msunyr$ and showed that the galaxies formed a BCG within $\sim300\,\Myr$, which actually predicted the \cite{deGraff2024} observation.  The \pkg{FOREVER22} simulations \citep{Yajima2021} were able to reproduce the star formation histories of starbursting protoclusters, but not the short depletion timescales in SPT2349-56.  Like \pkg{The Manhattan Suite}, \pkg{FOREVER22} is a suite of hydrodynamical cosmological simulations that focuses on the most massive objects at $z = 2$.  They have two relevant sets of simulations, protocluster regions and BCG regions.  Their protocluster regions are equivalent to my definition (i.e. roughly, $3\,\Rvir$ centered on the most massive halo) and their BCG regions are zoom-ins simulated out to $\Rvir$.  Their resolution is an order-of-magnitude better than \pkg{The Manhattan Suite} in the protocluster regions at $m_\mathrm{gas,part} \sim 5\times10^{6}\,\Msun$, and their BCG runs have $m_\mathrm{gas,part} \sim 5\times10^{5}\,\Msun$.  However, in both cases they only have $10$ objects and their parent volume is roughly $V\sim(740\,\cMpc)^3$.  My parent volume is $V\sim(1500\,\cMpc)^3$ and, therefore, has roughly $8$ times as many starbursting protocluster regions and I simulate $100$ of the most massive systems at $z = 2$.

\cite{Yajima2021} noted similar trends to what I describe in this work, that the SFRD peaks earlier than the field (i.e. \citealt{Madau2014}).  However, their explanation for the downturn is that stellar feedback and consumption dictate the shape of the SFRD in these systems.  That is counter to the picture that I present in Section~\ref{sec:galaxy_quenching}, and to the general idea that galaxies in massive halos are quenched primarily by AGN.  They supported their result with evidence from a control simulation with no AGN feedback, and showed that activating AGN only contributed a factor of two to the overall suppression.  However, it is important to note that (a) their halos are less massive (and fewer), (b) their SFRD peaks at lower values than in \pkg{The Manhattan Suite}, and (c) their suppression of the peak is less than in \pkg{The Manhattan Suite}.  In my opinion, that indicates that the AGN feedback is weaker than required and is evidenced through their AGN jet velocity of $v_\mathrm{jet} = 3000\,\mathrm{km}\,\mathrm{s}^{-1}$ whereas the \pkg{Simba} model has a maximum jet velocity of $v_\mathrm{jet}\sim7000\,\mathrm{km}\,\mathrm{s}^{-1}$ and was calibrated to reproduce the quenched population in massive galaxies at $z = 0$ \citep{Dave2019}.  

A similar study was presented in \cite{Remus2023} but they were more focused on finding something like the SPT2349-56 system, and did not run focused, zoom-in simulations.  In particular, they used \textit{Box 2b} of the \pkg{Magneticum Pathfinder} simulations (Dolag et al., in prep.), which is a $V = (940\,\cMpc)^3$ hydrodynamical cosmological simulation with particle resolution $m_\mathrm{gas} \sim 2\times10^{8}\,\Msun$.  That simulation has roughly $10\times$ lower resolution than \pkg{The Manhattan Suite}, and $100\times$ lower resolution than the \pkg{FOREVER22} simulations.  Similar to \pkg{FOREVER22}, the box volume is smaller than \pkg{The Manhattan Suite} and, therefore, contains roughly $4\times$ less massive systems.  However, they boast the ability to follow all of the cosmological gas flows given they have the entire volume simulated.  There are two important results from \cite{Remus2023}: (a) extreme protocluster regions do not necessarily end up as the most massive objects at $z = 0$, and (ii) the richness of the protocluster regions is a poor indicator of $z = 0$ cluster mass.  Unfortunately, they were unable to match the short depletion timescales in SPT2349-56, contrary to the results that I present in my work.  The reason for this difference is that their box size is too small to have the rarest fluctuations, or a representative sample of galaxy clusters at $z = 2$. Additionally, their resolution is poor and, as we showed in \cite{Lim2020}, increasing the resolution increases in the SFRs of galaxies --- exactly at the resolution transition regime between $10^7\,\Msun$ and $10^8\,\Msun$ per baryonic particle.  Comparatively, I show in this work that a simpler explanation is that SPT2349-56 is simply an outlier in the distribution of rare, extreme progenitors of $z = 2$ galaxy clusters.

Obviously, my results indicate that having a large box size is necessary to capture extreme starbursting regions in the early Universe.  Recently, the \pkg{FLAMINGO} large-volume, cosmological simulation suite was released that boasts volumes of $V\sim(1000\,\cMpc)^3$ and $V\sim(2.8\,\cGpc)^3$ at $m_\mathrm{gas} \sim7\times10^8\,\Msun$ and $m_\mathrm{gas}\sim7\times10^9\,\Msun$, respectively \citep{Schaye2023}.  The first is slightly bigger than \pkg{Magneticum} \textit{Box 2b}, but with a factor of two better mass resolution.  However, the mass resolution is a factor of two smaller than \pkg{The Manhattan Suite}, and the volume is smaller by a factor of three, so one would expect $3\times$ less massive structures at high redshift.  The second box is enormous, and should have (at most) $8\times$ as many massive halos as \pkg{The Manhattan Suite}.  However, the mass resolution is $50\times$ worse than \pkg{The Manhattan Suite}.  \cite{Lim2024} give an excellent analysis on the evolution of protocluster regions identified as Coma-like, Virgo-like, and Fornax-like in the \pkg{FLAMINGO} suite.  They clearly showed (their Figure 13) what I discuss above, that the box size is very important to capturing the overall SFRD history of protocluster regions at high redshift.  In particular, their SFRD from the \pkg{FLAMINGO} suite is unable to match the observations above $z > 2$.  That could be either because (a) their sub-grid feedback models are not calibrated in this extreme region or (b) their resolution is still insufficient.  

For (a), it is unlikely that the sub-grid prescriptions are the reason why the galaxies that I discuss are underrepresented. The majority of large-volume cosmological hydrodynamics simulations are unable to match the observed quenched number density of galaxies at $z > 3$ (\citealt{Merlin2019, Valentino2023}; although, see  \citealt{Hough2023}). If the simulations are unable to match the quenched densities, that implies that their simulated galaxy SFRs are too high at high redshift. If that is true, and they cannot reproduce the systems that I discuss in this work, then increasing the SFRs of galaxies must not be the correct solution.  Increasing the SFRs would only cause the simulated galaxy populations to be further from the observed quenched number densities.  All sub-grid models for star formation in large-volume simulations use a prescription similar to equation~\ref{eq:h2_star_formation_rate}, where the SFR scales with the gas density.  Therefore, I conclude that the only way to find systems such as those I discuss here is to find systems that have a higher central gas density --- massive halos at high redshift.

For (b), we showed in \cite{Lim2020} that the resolution plays a key factor in high SFRs.  Combining the result from \cite{Lim2020} and the result from \cite{Lim2024} on the box size, \pkg{The Manhattan Suite} exists in an excellent section of parameter space where there is a sufficient sample of overdense structures and also sufficient resolution to have representative SFRs.

The thread through the extremely massive (sometimes quenched) galaxies observed from $z = 2$ to $z = 8$, the extremely starbursting protoclusters, and the galaxy clusters at $z = 2$ is supported by my results, yet was not uncovered in the above-discussed simulations.  The reason is, as I stated, the box volume, resolution, and sample selection must capture sufficient systems with number densities $n\sim6\times10^{-8}\,\cMpc^{-3}$, or there will be bias in any interpretations.  For that reason, I intentionally neglect detailed discussion on other galaxy cluster simulation suites that construct their samples at $z = 0$.  Suites such as \pkg{Hydrangea} ($24$ zoom-ins; \citealt{Bahe2017}), \pkg{The 300} ($324$ zoom-ins; \citealt{Cui2018, Cui2022}), \pkg{FABLE} ($6$ zoom-ins; \citealt{Henden2018}), and \pkg{TNG-Cluster} ($356$ zoom-ins; \citealt{Nelson2024}) have a general approach of sampling $N$ objects in the mass scale range $\Mvir \gtrsim 10^{13}\,\Msun$ at $z = 0$.  Individually, they are distinct in their sampling techniques, but generally their samples do not properly sample the assembly histories of the galaxy clusters themselves (directly).  For example, \pkg{TNG-Cluster} has hundreds of $\Mvir > 10^{15}\,\Msun$ galaxy clusters at $z = 0$ but we know from \cite{Rennehan2020}, \cite{Remus2023}, and \cite{Lim2024} that the halo mass at $z = 0$ is not a good indicator of the extremely starbursting protoclusters that we observe in the Universe. 

Indeed, that is the most important result in this work.  From my intentionally-biased sample, I posit that the observed massive (sometimes quenched) galaxies at $z \gtrsim 3$ are in fact proto-BCGs of systems that become galaxy clusters at $z \sim 2$.  They will most likely not become high mass galaxy clusters at $z = 0$ \citep{Remus2023}, and they will get lost in the population statistics of clusters with $10^{14}\,\Msun < \Mvir|_\mathrm{z=0} < 10^{15}\,\Msun$.  Contemporary simulation suites miss the assembly histories of these systems, since their samples are not constructed to include the rarest of galaxies.  In future work, I will look for any indicators in $z = 0$ clusters that show whether they had a rapid growth phase at high redshift, followed by weak growth.

\section{Conclusions}
\label{sec:conclusions}

Observational advances have allowed for the detection and characterization of extreme objects at high redshift, such as extremely massive (sometimes quenched) galaxies at $z \gtrsim 3$, highly star-forming protoclusters at $z \gtrsim 3$, fully-assembled galaxy clusters at $z \sim 2$.  These systems push our understanding to the boundaries of what is possible within the $\Lambda\mathrm{CDM}$ framework, and challenge our theoretical models of galaxy evolution.  In this work, I present \pkg{The Manhattan Suite} --- a suite of $100$ focused zoom-in simulations of massive galaxy clusters ($\Mvir > 10^{14}\,\Msun$) selected at $z = 2$. I use the \pkg{Simba} galaxy evolution model as a base for my study \citep{Dave2019}.

I show that my simulated galaxy cluster population (including the BCG) reproduces the properties of the observed galaxy clusters XLSSC122 \citep{Willis2020} and JKCS041 \citep{Andreon2009}, that I use as case studies.  Specifically, \pkg{The Manhattan Suite} reproduces the quenched fraction all massive galaxies within the core and the mass of the BCG.  However, there is tension between the effective radii of the simulated BCGs compared to the observations, as well as a slight tension in the number of massive galaxies within the galaxy cluster cores.  I posit that both tensions could be the result of a timing issue of star formation and feedback acting in concert, similar to the arguments in \cite{RyanJoung2009}.

In Section~\ref{sec:galaxy_sfrs} I show that the protocluster regions have enhanced SFRD compared to the field observations in \cite{Madau2014}.  Not only are the protocluster region SFRD enhanced, but they also peak $\sim3\,\Gyr$ earlier than the field.  I suggest that this is \textit{accelerated galaxy evolution} due to the extreme overdensities of my systems, which have extremely short dynamical timescales.  The extremely short dynamical times lead to high gas densities and, therefore, high SFRs and low depletion times (see Fig.~\ref{fig:galaxy_evolution_fgas_tdep}).

In particular, the individual $100$ BCGs mostly quench between $z \sim 5$ and $z = 2$ when their sSFRs drop, while having mean stellar masses $\log_\mathrm{10} \overline{\Mstar} = 9.91, 10.9, 11.6, 12.0$ at $z = 8, 6, 4, 2$, respectively.  Simultaneously, their central SMBHs continue to grow.  In particular, the mean central SMBH masses are $\log_\mathrm{10} \overline{M}_\mathrm{BH} = 6.00, 8.21, 8.94, 10.1$ at $z = 8, 6, 4, 2$, respectively.  I show that the typical quasar and kinetic jet luminosities cross the typical core ($\rfive$) cooling luminosity at roughly $z \sim 5$ causing the decline of star formation at $z = 4$.  Past that point, the kinetic jet increases in power while the quasar power decreases.  I predict that most high redshift proto-BCGs should typically host a powerful AGN at $z \sim 4$ with $L_\mathrm{AGN} \sim 10^{45}\,\ergs$ with signatures of large-scale jet activity.  At higher redshifts, I predict that lower luminosity AGN should exist at $L_\mathrm{AGN} \gtrsim 10^{42}\,\ergs$ beginning at $z \sim 7$. 

One of the most important results of this work is the connection between extremely massive (sometimes quenched) galaxies at high redshift, extreme protoclusters like SPT2349-56, and galaxy clusters such as XLSSC122 and JKCS041.  I show in Section~\ref{sec:galaxy_populations} that \pkg{The Manhattan Suite} population reproduces both the the expected sSFRs and depletion times in SPT2349-56 at $z\sim4$ (cf. \citealt{Remus2023}).  The descendents of the SPT2349-56-like objects then go on to reproduce the high redshift galaxy clusters that have been observed at $z = 2$.  In Section~\ref{sec:galaxy_monsters}, I show that the evolution of stellar masses, sSFRs, effective radii, and stellar surface densities of the observed massive (sometimes quenched) galaxies at high redshift live within the most massive progenitor tracks of the BCGs of my galaxy clusters at $z = 2$.  Therefore, I draw the conclusion that the extremely massive galaxies at high redshift, highly star forming protoclusters at $z \gtrsim 3$, and the galaxy clusters like XLSSC122 and JKCS041 at $z = 2$ are all intimately linked.

I also argue in Section~\ref{sec:discussion} that if a simulation sample does not adequately sample halo assembly times, the link between these extreme objects can be obfuscated.  The evolutionary tracks of the most massive progenitors of my BCGs at $z = 2$ act as a predictive maximum upper envelope for the galaxy stellar masses, sSFRs, effective radii, and stellar surface densities of high redshift, massive galaxies since I construct my sample to have a biased-early ($z\sim2$) assembly history.  Galaxies that are very massive at lower redshift may live in the progenitor histories of galaxy clusters that assemble below my selection redshift, $z < 2$.  That echoes our prediction in \cite{Rennehan2020}, where there is a assembly history distribution of massive BCGs at all redshifts, in the cores of galaxy clusters.

There are many avenues of future research into these interesting systems.  The most important next step is, in my opinion, understanding how the distribution of halo assembly times influences the observed distribution of massive quenched galaxies. The scope of my simulation suite is unable to conclusively answer the question given the small, highly focused volumes and, therefore, our theoretical models remain in tension with a subset of the high redshift, massive quenched galaxies.  I am hopeful that the issue will be resolved by studying galaxy populations as a function of overdensity, especially given the success of approaches such as \pkg{FLARES} \citep{Lovell2021}.

\section*{Acknowledgements}

The simulations presented in this work were run on the Flatiron Institute's research computing facilities (the Iron compute cluster), supported by the Simons Foundation.  I am supported by the Simons Foundation.  I thank Romeel Dav\'{e} for providing the \pkg{Simba} code for this project and for helpful discussions, and Ulrich P. Steinwandel for the parent N-body simulation. I additionally thank Arif Babul, Romeel Dav\'{e}, Rachel S. Somerville, Greg L. Bryan, Scott C. Chapman, Elena Martínez Hernández, Romain Teyssier, Rachel K. Cochrane, Christopher C. Hayward, Ulrich P. Steinwandel, and Asya Rennehan for insightful discussions during the course of this research. Finally, I thank the anonymous referee for improving this manuscript.

\facility{Rusty (Flatiron Institute)}

\software{Our analysis was performed using the Python programming language.  The following packages were used throughout the analysis: \pkg{h5py} \citep{Collette2013}, \pkg{numpy} \citep{Harris2020}, \pkg{scipy} \citep{Virtanen2020}, \pkg{yt} \citep{Turk2011}, and \pkg{matplotlib} \citep{Hunter2007}.  Prototyping of the analysis scripts was performed in the IPython environment \citep{Perez2007}. I use \pkg{MUSIC} to generate initial conditions \citep{Hahn2011}, \pkg{Gadget-4} to run the pre-flight simulations \citep{Springel2021}, \pkg{Rockstar} to find halos in my simulation \citep{Behroozi2013}, \pkg{GIZMO} for my hydrodynamics simulation \citep{Hopkins2015a}, and the \pkg{Grackle} cooling library \citep{Smith2017}.}





\clearpage
\appendix 

\section{Power law fits} \label{app:power_law_fits}

\begin{deluxetable}{lcccc}[h]
\tablecaption{Power-law fit parameters from equation~\ref{eq:power_law_fit}.  $\mu_\mathrm{a}$ is the mean slope across all power law fits to each individual cluster, and $\sigma_\mathrm{a}$ is the standard deviation of the distribution.  The mean and standard deviations come from a Gaussian fit to the entire population of fit parameters.  Similarly, $\mu_\mathrm{b}$ is the mean intercept in equation~\ref{eq:power_law_fit} and $\sigma_\mathrm{b}$ is the standard deviation.  Where possible, I have included the observed power law fits with the associated errors. 
 \label{tab:power_law_parameters}}
\tablecolumns{6}
\tablewidth{0pt}
\tablehead{
\colhead{Population quantity} & \colhead{$\mu_\mathrm{a}$} & \colhead{$\sigma_\mathrm{a}$} &
\colhead{$\mu_\mathrm{b}$} & \colhead{$\sigma_\mathrm{b}$}
}
\startdata
Stellar mass function & -0.263 & 0.0936 & 3.68 &  1.34 \\
XLSSC122 & -0.585 & n/a & 7.4 & n/a \\
\hline
Effective radii & 0.214 & 0.0386 & -1.9 & 0.371 \\
XLSSC122 & 0.613 & n/a & -6.1 & n/a \\
3DHST & 0.747 & n/a & -7.74 & n/a \\
\hline
Black hole masses & 1.88 & 0.235 & -12.9 & 2.59 \\
\citealt{Kormendy2013} & 1.17 & 0.08 & -4.18 & 0.1 \\
\citealt{Bentz2018} & 1.84 & 0.25 & -10.7 & 0.265 \\
\enddata
\end{deluxetable}

\begin{deluxetable}{lcccc}[h]
\tablecaption{Fit parameters for equation~\ref{eq:madau_dickinson}, comparing the field observations (first row) and the typical simulated protocluster region (second row). \label{tab:sfrd_parameters}}
\tablecolumns{6}
\tablewidth{0pt}
\tablehead{
\colhead{Data set} & \colhead{a} & \colhead{b} &
\colhead{c} & \colhead{d}
}
\startdata
\cite{Madau2014} & 0.015 & 2.7 & 2.9 & 5.6 \\
\pkg{The Manhattan Suite} & 0.0026 & 5.5 & 5.0 & 11.5 \\
\enddata
\end{deluxetable}

Table~\ref{tab:power_law_parameters} shows the power law fit parameters to the population of simulated galaxy clusters and brightest cluster galaxies at $z = 2$.  Where possible, I include the observed values for the scaling relationships and the associated error.  Unfortunately, for 3DHST and XLSSC122, \cite{Noordeh2021} do not provide their fit parameters and errors.  For the stellar mass function, I fit a power law with slope $a$ and intercept $b$ to the histogram of massive galaxies ($>10^{10}\,\Msun$) within $400\,\kpc$ of the cluster center.  To find the parameter population mean and standard deviation, I fit a Gaussian to all of the parameters $\theta_{ij} = (a_{i}, b_{i})$ for each galaxy cluster $i$,

\begin{equation}
    \label{eq:gaussian}
    f(\theta) = Ae^{-(\theta - \mu_\mathrm{\theta})^2 / (2 \sigma_\mathrm{\theta}^2)},
\end{equation}

\noindent where $\theta$ is the parameter in the power law, $\mu_\mathrm{\theta}$ is the population mean, and $\sigma_\mathrm{\theta}$ is the population standard deviation.  When a Gaussian fit fails, I use the mean and standard deviation of the distribution.  For the effective radii and the black hole masses, I apply the same procedure except that I use the brightest cluster galaxies only, rather than all of the massive galaxies within the core.

Table~\ref{tab:sfrd_parameters} shows the mean SFRD parameter values for the entire population of protocluster regions.  I first fit equation~\ref{eq:madau_dickinson} to each protocluster region, as a function of redshift.  Then, I find the distributions of the parameters $\theta_{ij} = (a_i,b_i,c_i,d_i)$.  I fit each distribution with a Gaussian function, as in equation~\ref{eq:gaussian}.  I do not show the standard deviations for the SFRD, as I am not intending to show consistency between the observed field relationship and the simulated relationship.  The useful information is in the rough differences between the parameters.

\bibliography{main}{}

\begin{thebibliography}{}
\expandafter\ifx\csname natexlab\endcsname\relax\def\natexlab#1{#1}\fi
\providecommand{\url}[1]{\href{#1}{#1}}
\providecommand{\dodoi}[1]{doi:~\href{http://doi.org/#1}{\nolinkurl{#1}}}
\providecommand{\doeprint}[1]{\href{http://ascl.net/#1}{\nolinkurl{http://ascl.net/#1}}}
\providecommand{\doarXiv}[1]{\href{https://arxiv.org/abs/#1}{\nolinkurl{https://arxiv.org/abs/#1}}}

\bibitem[{{\'{A}}lvarez-M{\'{a}}rquez {et~al.}(2023){\'{A}}lvarez-M{\'{a}}rquez, {Crespo G{\'{o}}mez}, Colina, Neeleman, Walter, Labiano, P{\'{e}}rez-Gonz{\'{a}}lez, Bik, Noorgaard-Nielsen, Ostlin, Wright, Alonso-Herrero, Azollini, Caputi, Eckart, {Le F{\`{e}}vre}, Garc{\'{i}}a-Mar{\'{i}}n, Greve, Hjorth, Ilbert, Kendrew, Pye, Tikkanen, Topinka, van~der Werf, Ward, van Dishoeck, G{\"{u}}del, Henning, Lagage, Ray, \& Waelkens}]{AlvarezMarquez2023}
{\'{A}}lvarez-M{\'{a}}rquez, J., {Crespo G{\'{o}}mez}, A., Colina, L., {et~al.} 2023, Astronomy {\&} Astrophysics, 671, A105, \dodoi{10.1051/0004-6361/202245400}

\bibitem[{Andreon {et~al.}(2009)Andreon, Maughan, Trinchieri, \& Kurk}]{Andreon2009}
Andreon, S., Maughan, B., Trinchieri, G., \& Kurk, J. 2009, Astronomy {\&} Astrophysics, 507, 147, \dodoi{10.1051/0004-6361/200912299}

\bibitem[{Andreon {et~al.}(2014)Andreon, Newman, Trinchieri, Raichoor, Ellis, \& Treu}]{Andreon2014}
Andreon, S., Newman, A.~B., Trinchieri, G., {et~al.} 2014, Astronomy {\&} Astrophysics, 565, A120, \dodoi{10.1051/0004-6361/201323077}

\bibitem[{Andreon {et~al.}(2021)Andreon, Romero, Castagna, Ragagnin, Devlin, Dicker, Mason, Mroczkowski, Sarazin, Sievers, \& Stanchfield}]{Andreon2021}
Andreon, S., Romero, C., Castagna, F., {et~al.} 2021, Monthly Notices of the Royal Astronomical Society, 505, 5896, \dodoi{10.1093/mnras/stab1639}

\bibitem[{Angl{\'{e}}s-Alc{\'{a}}zar {et~al.}(2017)Angl{\'{e}}s-Alc{\'{a}}zar, Faucher-Gigu{\`{e}}re, Kere{\v{s}}, Hopkins, Quataert, \& Murray}]{Angles2017}
Angl{\'{e}}s-Alc{\'{a}}zar, D., Faucher-Gigu{\`{e}}re, C.-A., Kere{\v{s}}, D., {et~al.} 2017, Monthly Notices of the Royal Astronomical Society, 470, 4698, \dodoi{10.1093/mnras/stx1517}

\bibitem[{Bahcall(1977)}]{Bahcall1977}
Bahcall, N.~A. 1977, Annual Review of Astronomy and Astrophysics, 15, 505, \dodoi{10.1146/annurev.aa.15.090177.002445}

\bibitem[{Bah{\'{e}} {et~al.}(2017)Bah{\'{e}}, Barnes, {Dalla Vecchia}, Kay, White, McCarthy, Schaye, Bower, Crain, Theuns, Jenkins, McGee, Schaller, Thomas, \& Trayford}]{Bahe2017}
Bah{\'{e}}, Y.~M., Barnes, D.~J., {Dalla Vecchia}, C., {et~al.} 2017, Monthly Notices of the Royal Astronomical Society, 470, 4186, \dodoi{10.1093/mnras/stx1403}

\bibitem[{Behroozi {et~al.}(2013{\natexlab{a}})Behroozi, Wechsler, \& Conroy}]{Behroozi2013a}
Behroozi, P.~S., Wechsler, R.~H., \& Conroy, C. 2013{\natexlab{a}}, The Astrophysical Journal, 770, 57, \dodoi{10.1088/0004-637X/770/1/57}

\bibitem[{Behroozi {et~al.}(2013{\natexlab{b}})Behroozi, Wechsler, \& Wu}]{Behroozi2013}
Behroozi, P.~S., Wechsler, R.~H., \& Wu, H.-Y. 2013{\natexlab{b}}, The Astrophysical Journal, 762, 109, \dodoi{10.1088/0004-637X/762/2/109}

\bibitem[{Bentz \& Manne-Nicholas(2018)}]{Bentz2018}
Bentz, M.~C., \& Manne-Nicholas, E. 2018, The Astrophysical Journal, 864, 146, \dodoi{10.3847/1538-4357/aad808}

\bibitem[{Bezanson {et~al.}(2022)Bezanson, Labbe, Whitaker, Leja, Price, Franx, Brammer, Marchesini, Zitrin, Wang, Weaver, Furtak, Atek, Coe, Cutler, Dayal, van Dokkum, Feldmann, Schreiber, Fujimoto, Geha, Glazebrook, de~Graaff, Greene, Juneau, Kassin, Kriek, Khullar, Maseda, Mowla, Muzzin, Nanayakkara, Nelson, Oesch, Pacifici, Pan, Papovich, Setton, Shapley, Smit, Stefanon, Taylor, \& Williams}]{Bezanson2022}
Bezanson, R., Labbe, I., Whitaker, K.~E., {et~al.} 2022, arXiv preprint arXiv:2212.04026.
\newblock \doarXiv{2212.04026}

\bibitem[{Bhowmick {et~al.}(2020)Bhowmick, Somerville, {Di Matteo}, Wilkins, Feng, \& Tenneti}]{Bhowmick2020}
Bhowmick, A.~K., Somerville, R.~S., {Di Matteo}, T., {et~al.} 2020, Monthly Notices of the Royal Astronomical Society, 496, 754, \dodoi{10.1093/mnras/staa1605}

\bibitem[{Bondi(1952)}]{Bondi1952}
Bondi, H. 1952, Monthly Notices of the Royal Astronomical Society, 112, 195, \dodoi{10.1093/mnras/112.2.195}

\bibitem[{Boylan-Kolchin(2023)}]{BoylanKolchin2023}
Boylan-Kolchin, M. 2023, Nature Astronomy, 7, 731, \dodoi{10.1038/s41550-023-01937-7}

\bibitem[{Bruzual \& Charlot(2003)}]{Bruzual2003}
Bruzual, G., \& Charlot, S. 2003, Monthly Notices of the Royal Astronomical Society, 344, 1000, \dodoi{10.1046/j.1365-8711.2003.06897.x}

\bibitem[{Bryan \& Norman(1998)}]{Bryan1998}
Bryan, G.~L., \& Norman, M.~L. 1998, The Astrophysical Journal, 495, 80, \dodoi{10.1086/305262}

\bibitem[{Carnall {et~al.}(2024)Carnall, Cullen, McLure, McLeod, Begley, Donnan, Dunlop, Shapley, Rowlands, Almaini, Arellano-C{\'{o}}rdova, Barrufet, Cimatti, Ellis, Grogin, Hamadouche, Illingworth, Koekemoer, Leung, Lovell, P{\'{e}}rez-Gonz{\'{a}}lez, Santini, Stanton, \& Wild}]{Carnall2024}
Carnall, A.~C., Cullen, F., McLure, R.~J., {et~al.} 2024, arXiv preprint arXiv:2405.02242.
\newblock \doarXiv{2405.02242}

\bibitem[{Casey {et~al.}(2023)Casey, Kartaltepe, Drakos, Franco, Harish, Paquereau, Ilbert, Rose, Cox, Nightingale, Robertson, Silverman, Koekemoer, Massey, McCracken, Rhodes, Akins, Allen, Amvrosiadis, Arango-Toro, Bagley, Bongiorno, Capak, Champagne, Chartab, {Ch{\'{a}}vez Ortiz}, Chworowsky, Cooke, Cooper, Darvish, Ding, Faisst, Finkelstein, Fujimoto, Gentile, Gillman, Gould, Gozaliasl, Hayward, He, Hemmati, Hirschmann, Jahnke, Jin, Khostovan, Kokorev, Lambrides, Laigle, Larson, Leung, Liu, Liaudat, Long, Magdis, Mahler, Mainieri, Manning, Maraston, Martin, McCleary, McKinney, McPartland, Mobasher, Pattnaik, Renzini, Rich, Sanders, Sattari, Scognamiglio, Scoville, Sheth, Shuntov, Sparre, Suzuki, Talia, Toft, Trakhtenbrot, Urry, Valentino, Vanderhoof, Vardoulaki, Weaver, Whitaker, Wilkins, Yang, \& Zavala}]{Casey2023}
Casey, C.~M., Kartaltepe, J.~S., Drakos, N.~E., {et~al.} 2023, The Astrophysical Journal, 954, 31, \dodoi{10.3847/1538-4357/acc2bc}

\bibitem[{Chabrier(2003)}]{Chabrier2003}
Chabrier, G. 2003, Publications of the Astronomical Society of the Pacific, 115, 763, \dodoi{10.1086/376392}

\bibitem[{Chapman {et~al.}(2024)Chapman, Hill, Aravena, Archipley, Babul, Burgoyne, Canning, Deane, {De Breuck}, Gonzalez, Hayward, Kim, Malkan, Marrone, McIntyre, Murphy, Pass, Perry, Phadke, Rennehan, Reuter, Rotermund, Scott, Seymour, Solimano, Spilker, Stark, Sulzenauer, Tothill, Vieira, Vizgan, Wang, \& Weiss}]{Chapman2024}
Chapman, S.~C., Hill, R., Aravena, M., {et~al.} 2024, The Astrophysical Journal, 961, 120, \dodoi{10.3847/1538-4357/ad0b77}

\bibitem[{Chiang {et~al.}(2017)Chiang, Overzier, Gebhardt, \& Henriques}]{Chiang2017}
Chiang, Y.-K., Overzier, R.~A., Gebhardt, K., \& Henriques, B. 2017, The Astrophysical Journal, 844, L23, \dodoi{10.3847/2041-8213/aa7e7b}

\bibitem[{Choi {et~al.}(2012)Choi, Ostriker, Naab, \& Johansson}]{Choi2012a}
Choi, E., Ostriker, J.~P., Naab, T., \& Johansson, P.~H. 2012, The Astrophysical Journal, 754, 125, \dodoi{10.1088/0004-637X/754/2/125}

\bibitem[{Choi {et~al.}(2018)Choi, Somerville, Ostriker, Naab, \& Hirschmann}]{Choi2018}
Choi, E., Somerville, R.~S., Ostriker, J.~P., Naab, T., \& Hirschmann, M. 2018, The Astrophysical Journal, 866, 91, \dodoi{10.3847/1538-4357/aae076}

\bibitem[{Chworowsky {et~al.}(2023)Chworowsky, Finkelstein, Boylan-Kolchin, McGrath, Iyer, Papovich, Dickinson, Taylor, Yung, Haro, Bagley, Backhaus, Bhatawdekar, Cheng, Cleri, Cole, Cooper, Costantin, Dekel, Franco, Fujimoto, Hayward, Holwerda, Huertas-Company, Hirschmann, Hutchison, Koekemoer, Larson, Li, Long, Lucas, Pirzkal, Rodighiero, Somerville, Vanderhoof, de~la Vega, Wilkins, Yang, \& Zavala}]{Chworowsky2023b}
Chworowsky, K., Finkelstein, S.~L., Boylan-Kolchin, M., {et~al.} 2023, arXiv preprint arXiv:2311.14804.
\newblock \doarXiv{2311.14804}

\bibitem[{Collette(2013)}]{Collette2013}
Collette, A. 2013, {Python and HDF5} (O'Reilly), 152

\bibitem[{Collins {et~al.}(2009)Collins, Stott, Hilton, Kay, Stanford, Davidson, Hosmer, Hoyle, Liddle, Lloyd-Davies, Mann, Mehrtens, Miller, Nichol, Romer, Sahl{\'{e}}n, Viana, \& West}]{Collins2009}
Collins, C.~A., Stott, J.~P., Hilton, M., {et~al.} 2009, Nature, 458, 603, \dodoi{10.1038/nature07865}

\bibitem[{Conroy {et~al.}(2015)Conroy, van Dokkum, \& Kravtsov}]{Conroy2015a}
Conroy, C., van Dokkum, P.~G., \& Kravtsov, A. 2015, The Astrophysical Journal, 803, 77, \dodoi{10.1088/0004-637X/803/2/77}

\bibitem[{Cui {et~al.}(2018)Cui, Knebe, Yepes, Pearce, Power, Dave, Arth, Borgani, Dolag, Elahi, Mostoghiu, Murante, Rasia, Stoppacher, Vega-Ferrero, Wang, Yang, Benson, Cora, Croton, Sinha, Stevens, Vega-Mart{\'{i}}nez, Arthur, Baldi, Ca{\~{n}}as, Cialone, Cunnama, {De Petris}, Durando, Ettori, Gottl{\"{o}}ber, Nuza, Old, Pilipenko, Sorce, \& Welker}]{Cui2018}
Cui, W., Knebe, A., Yepes, G., {et~al.} 2018, Monthly Notices of the Royal Astronomical Society, 480, 2898, \dodoi{10.1093/mnras/sty2111}

\bibitem[{Cui {et~al.}(2022)Cui, Dave, Knebe, Rasia, Gray, Pearce, Power, Yepes, Anbajagane, Ceverino, Contreras-Santos, de~Andres, {De Petris}, Ettori, Haggar, Li, Wang, Yang, Borgani, Dolag, Zu, Kuchner, Ca{\~{n}}as, Ferragamo, \& Gianfagna}]{Cui2022}
Cui, W., Dave, R., Knebe, A., {et~al.} 2022, Monthly Notices of the Royal Astronomical Society, 514, 977, \dodoi{10.1093/mnras/stac1402}

\bibitem[{Curtis-Lake {et~al.}(2023)Curtis-Lake, Carniani, Cameron, Charlot, Jakobsen, Maiolino, Bunker, Witstok, Smit, Chevallard, Willott, Ferruit, Arribas, Bonaventura, Curti, D'Eugenio, Franx, Giardino, Looser, L{\"{u}}tzgendorf, Maseda, Rawle, Rix, {Rodr{\'{i}}guez del Pino}, {\"{U}}bler, Sirianni, Dressler, Egami, Eisenstein, Endsley, Hainline, Hausen, Johnson, Rieke, Robertson, Shivaei, Stark, Tacchella, Williams, Willmer, Bhatawdekar, Bowler, Boyett, Chen, de~Graaff, Helton, Hviding, Jones, Kumari, Lyu, Nelson, Perna, Sandles, Saxena, Suess, Sun, Topping, Wallace, \& Whitler}]{CurtisLake2023}
Curtis-Lake, E., Carniani, S., Cameron, A., {et~al.} 2023, Nature Astronomy, 7, 622, \dodoi{10.1038/s41550-023-01918-w}

\bibitem[{Dav{\'{e}} {et~al.}(2019)Dav{\'{e}}, Angl{\'{e}}s-Alc{\'{a}}zar, Narayanan, Li, Rafieferantsoa, \& Appleby}]{Dave2019}
Dav{\'{e}}, R., Angl{\'{e}}s-Alc{\'{a}}zar, D., Narayanan, D., {et~al.} 2019, Monthly Notices of the Royal Astronomical Society, 486, 2827, \dodoi{10.1093/mnras/stz937}

\bibitem[{Dav{\'{e}} {et~al.}(2016)Dav{\'{e}}, Thompson, \& Hopkins}]{Dave2016c}
Dav{\'{e}}, R., Thompson, R., \& Hopkins, P.~F. 2016, Monthly Notices of the Royal Astronomical Society, 462, 3265, \dodoi{10.1093/mnras/stw1862}

\bibitem[{de~Graaff {et~al.}(2022)de~Graaff, Trayford, Franx, Schaller, Schaye, \& van~der Wel}]{deGraaff2022}
de~Graaff, A., Trayford, J., Franx, M., {et~al.} 2022, Monthly Notices of the Royal Astronomical Society, 511, 2544, \dodoi{10.1093/mnras/stab3510}

\bibitem[{de~Graaff {et~al.}(2024)de~Graaff, Setton, Brammer, Cutler, Suess, Labbe, Leja, Weibel, Maseda, Whitaker, Bezanson, Boogaard, Cleri, {De Lucia}, Franx, Greene, Hirschmann, Matthee, McConachie, Naidu, Oesch, Price, Rix, Valentino, Wang, \& Williams}]{deGraff2024}
de~Graaff, A., Setton, D.~J., Brammer, G., {et~al.} 2024, arXiv preprint arXiv:2404.05683.
\newblock \doarXiv{2404.05683}

\bibitem[{de~Putter {et~al.}(2012)de~Putter, Wagner, Mena, Verde, \& Percival}]{dePutter2012}
de~Putter, R., Wagner, C., Mena, O., Verde, L., \& Percival, W.~J. 2012, Journal of Cosmology and Astroparticle Physics, 2012, 019, \dodoi{10.1088/1475-7516/2012/04/019}

\bibitem[{Dekel {et~al.}(2023)Dekel, Sarkar, Birnboim, Mandelker, \& Li}]{Dekel2023}
Dekel, A., Sarkar, K.~C., Birnboim, Y., Mandelker, N., \& Li, Z. 2023, Monthly Notices of the Royal Astronomical Society, 523, 3201, \dodoi{10.1093/mnras/stad1557}

\bibitem[{Dubois {et~al.}(2016)Dubois, Peirani, Pichon, Devriendt, Gavazzi, Welker, \& Volonteri}]{Dubois2016}
Dubois, Y., Peirani, S., Pichon, C., {et~al.} 2016, Monthly Notices of the Royal Astronomical Society, 463, 3948, \dodoi{10.1093/mnras/stw2265}

\bibitem[{Endsley {et~al.}(2023)Endsley, Stark, Whitler, Topping, Chen, Plat, Chisholm, \& Charlot}]{Endsley2023}
Endsley, R., Stark, D.~P., Whitler, L., {et~al.} 2023, Monthly Notices of the Royal Astronomical Society, 524, 2312, \dodoi{10.1093/mnras/stad1919}

\bibitem[{Faucher-Gigu{\`{e}}re(2018)}]{Faucher-Giguere2018}
Faucher-Gigu{\`{e}}re, C.-A. 2018, Monthly Notices of the Royal Astronomical Society, 473, 3717, \dodoi{10.1093/mnras/stx2595}

\bibitem[{Faucher-Gigu{\`{e}}re {et~al.}(2009)Faucher-Gigu{\`{e}}re, Lidz, Zaldarriaga, \& Hernquist}]{Faucher2009}
Faucher-Gigu{\`{e}}re, C.-A., Lidz, A., Zaldarriaga, M., \& Hernquist, L. 2009, The Astrophysical Journal, 703, 1416, \dodoi{10.1088/0004-637X/703/2/1416}

\bibitem[{Finkelstein {et~al.}(2013)Finkelstein, Papovich, Dickinson, Song, Tilvi, Koekemoer, Finkelstein, Mobasher, Ferguson, Giavalisco, Reddy, Ashby, Dekel, Fazio, Fontana, Grogin, Huang, Kocevski, Rafelski, Weiner, \& Willner}]{Finkelstein2013}
Finkelstein, S.~L., Papovich, C., Dickinson, M., {et~al.} 2013, Nature, 502, 524, \dodoi{10.1038/nature12657}

\bibitem[{Finkelstein {et~al.}(2023)Finkelstein, Bagley, Ferguson, Wilkins, Kartaltepe, Papovich, Yung, {Arrabal Haro}, Behroozi, Dickinson, Kocevski, Koekemoer, Larson, {Le Bail}, Morales, P{\'{e}}rez-Gonz{\'{a}}lez, Burgarella, Dav{\'{e}}, Hirschmann, Somerville, Wuyts, Bromm, Casey, Fontana, Fujimoto, Gardner, Giavalisco, Grazian, Grogin, Hathi, Hutchison, Jha, Jogee, Kewley, Kirkpatrick, Long, Lotz, Pentericci, Pierel, Pirzkal, Ravindranath, Ryan, Trump, Yang, Bhatawdekar, Bisigello, Buat, Calabr{\`{o}}, Castellano, Cleri, Cooper, Croton, Daddi, Dekel, Elbaz, Franco, Gawiser, Holwerda, Huertas-Company, Jaskot, Leung, Lucas, Mobasher, Pandya, Tacchella, Weiner, \& Zavala}]{Finkelstein2023}
Finkelstein, S.~L., Bagley, M.~B., Ferguson, H.~C., {et~al.} 2023, The Astrophysical Journal Letters, 946, L13, \dodoi{10.3847/2041-8213/acade4}

\bibitem[{Fiore {et~al.}(2017)Fiore, Feruglio, Shankar, Bischetti, Bongiorno, Brusa, Carniani, Cicone, Duras, Lamastra, Mainieri, Marconi, Menci, Maiolino, Piconcelli, Vietri, \& Zappacosta}]{Fiore2017}
Fiore, F., Feruglio, C., Shankar, F., {et~al.} 2017, Astronomy {\&} Astrophysics, 601, A143, \dodoi{10.1051/0004-6361/201629478}

\bibitem[{Fujimoto {et~al.}(2023)Fujimoto, {Arrabal Haro}, Dickinson, Finkelstein, Kartaltepe, Larson, Burgarella, Bagley, Behroozi, Chworowsky, Hirschmann, Trump, Wilkins, Yung, Koekemoer, Papovich, Pirzkal, Ferguson, Fontana, Grogin, Grazian, Kewley, Kocevski, Lotz, Pentericci, Ravindranath, Somerville, Wilkins, Amor{\'{i}}n, Backhaus, Calabr{\`{o}}, Casey, Cooper, Fern{\'{a}}ndez, Franco, Giavalisco, Hathi, Harish, Hutchison, Iyer, Jung, Lucas, \& Zavala}]{Fujimoto2023}
Fujimoto, S., {Arrabal Haro}, P., Dickinson, M., {et~al.} 2023, The Astrophysical Journal Letters, 949, L25, \dodoi{10.3847/2041-8213/acd2d9}

\bibitem[{Gaburov \& Nitadori(2011)}]{Gaburov2011}
Gaburov, E., \& Nitadori, K. 2011, Monthly Notices of the Royal Astronomical Society, 414, 129, \dodoi{10.1111/j.1365-2966.2011.18313.x}

\bibitem[{Gardner {et~al.}(2023)Gardner, Mather, Abbott, Abell, Abernathy, Abney, Abraham, Abraham, Abul-Huda, Acton, Adams, Adams, Adler, Adriaensen, Aguilar, Ahmed, Ahmed, Ahmed, Albat, Albert, Alberts, Aldridge, Allen, Allen, Altenburg, Altunc, Alvarez, {\'{A}}lvarez-M{\'{a}}rquez, de~Oliveira, Ambrose, Anandakrishnan, Andersen, Anderson, Anderson, Anderson, Anderson, Aprea, Archer, Arenberg, Argyriou, Arribas, Artigau, Arvai, Atcheson, Atkinson, Averbukh, Aymergen, Bacinski, Baggett, Bagnasco, Baker, Balzano, Banks, Baran, Barker, Barrett, Barringer, Barto, Bast, Baudoz, Baum, Beatty, Beaulieu, Bechtold, Beck, Beddard, Beichman, Bellagama, Bely, Berger, Bergeron, Bernier, Bertch, Beskow, Betz, Biagetti, Birkmann, Bjorklund, Blackwood, Blazek, Blossfeld, Bluth, Boccaletti, {Boegner Jr}, Bohlin, Boia, B{\"{o}}ker, Bonaventura, Bond, Bosley, Boucarut, Bouchet, Bouwman, Bower, Bowers, Bowers, Boyce, Boyer, Boyer, Boyer, Boyer, Bradley, Brady, Brandl, Brannen, Breda, Bremmer, Brennan, Bresnahan, Bright,
  Broiles, Bromenschenkel, Brooks, Brooks, Brown, Brown, Brown, Bruce, Bryson, Bujanda, Bullock, Bunker, Bureo, Burt, Bush, Bushouse, Bussman, Cabaud, Cale, Calhoon, Calvani, Canipe, Caputo, Cara, Carey, Case, Cesari, Cetorelli, Chance, Chandler, Chaney, Chapman, Charlot, Chayer, Cheezum, Chen, Chen, Cherinka, Chichester, Chilton, Chittiraibalan, Clampin, Clark, Clark, Clark, Claybrooks, Cleveland, Cohen, Cohen, Col{\'{o}}n, Coleman, Colina, Comber, Comeau, Comer, Reis, Connolly, Conroy, Contos, Contreras, Cook, Cooper, Cooper, Correia, Correnti, Cossou, Costanza, Coulais, Cox, Coyle, Cracraft, Crew, Curtis, Cusveller, Maciel, Dailey, Daugeron, Davidson, Davies, Davis, Davis, Day, de~Chambure, de~Jong, {De Marchi}, Dean, Decker, Delisa, Dell, Dellagatta, Dembinska, Demosthenes, Dencheva, Deneu, DePriest, Deschenes, Dethienne, Detre, Diaz, Dicken, DiFelice, Dillman, Disharoon, Dixon, Doggett, Dominguez, Donaldson, Doria-Warner, Santos, Doty, {Douglas, Jr}, Doyon, Dressler, Driggers, Driggers, Dunn, DuPrie,
  Dupuis, Durning, Dutta, Earl, Eccleston, Ecobichon, Egami, Ehrenwinkler, Eisenhamer, Eisenhower, Eisenstein, {El Hamel}, Elie, Elliott, Elliott, Engesser, Espinoza, Etienne, Etxaluze, Evans, Fabreguettes, Falcolini, Falini, Fatig, Feeney, Feinberg, Fels, Ferdous, Ferguson, Ferrarese, Ferreira, Ferruit, Ferry, Filippazzo, Firre, Fix, Flagey, Flanagan, Fleming, Florian, Flynn, Foiadelli, Fontaine, Fontanella, Forshay, Fortner, Fox, Framarini, Francisco, Franck, Franx, Franz, Friedman, Friend, Frost, Fu, Fullerton, Gaillard, Galkin, Gallagher, Galyer, {Garc{\'{i}}a Mar{\'{i}}n}, Gardner, Garland, Garrett, Gasman, G{\'{a}}sp{\'{a}}r, Gastaud, Gaudreau, Gauthier, Geers, Geithner, Gennaro, Gerber, Gereau, Giampaoli, Giardino, Gibbons, Gilbert, Gilman, Girard, Giuliano, Gkountis, Glasse, Glassmire, Glauser, Glazer, Goldberg, Golimowski, Gonzaga, Gordon, Gordon, Goudfrooij, Gough, Graham, Grau, Green, Greene, Greene, Greenfield, Greenhouse, Greve, Greville, Grimaldi, Groe, Groebner, Grumm, Grundy, G{\"{u}}del,
  Guillard, Guldalian, Gunn, Gurule, Gutman, Guy, Guyot, Hack, Haderlein, Hagan, Hagedorn, Hainline, Haley, Hami, Hamilton, Hammann, Hammel, Hanley, Hansen, Hardy, Harnisch, Harr, Harris, Hart, Hartig, Hasan, Hashim, Hashimoto, Haskins, Hawkins, Hayden, Hayden, Healy, Hecht, Heeg, Hejal, Helm, Hengemihle, Henning, Henry, Henry, Henshaw, Hernandez, Herrington, Heske, Hesman, Hickey, Hilbert, Hines, Hinz, Hirsch, Hitcho, Hodapp, Hodge, Hoffman, Holfeltz, Holler, Hoppa, Horner, Howard, Howard, Huber, Hunkeler, Hunter, Hunter, Hurd, Hurst, Hutchings, Hylan, Ignat, Illingworth, Irish, {Isaacs III}, {Jackson Jr}, Jaffe, Jahic, Jahromi, Jakobsen, James, James, James, Jamieson, Jandra, Jayawardhana, Jedrzejewski, Jeffers, Jensen, Joanne, Johns, Johnson, Johnson, Johnson, Johnson, Johnson, Johnson, Johnstone, Jollet, Jones, Jones, Jones, Jones, Jones, Jordan, Jordan, Jue, Jurkowski, Justis, Justtanont, Kaleida, Kalirai, Kalmanson, Kaltenegger, Kammerer, Kan, Kanarek, Kao, Karakla, Karl, Kassin, Kauffman, Kavanagh,
  Kelley, Kelly, Kendrew, Kennedy, Kenny, Keski-Kuha, Keyes, Khan, Kidwell, Kimble, King, King, Kinzel, Kirk, Kirkpatrick, Klaassen, Klingemann, Klintworth, Knapp, Knight, Knollenberg, Knutsen, Koehler, Koekemoer, Kofler, Kontson, Kovacs, Kozhurina-Platais, Krause, Kriss, Krist, Kristoffersen, Krogel, Krueger, Kulp, Kumari, Kwan, Kyprianou, Labador, Labiano, Lafreni{\`{e}}re, Lagage, Laidler, Laine, Laird, Lajoie, Lallo, Lam, LaMassa, Lambros, Lampenfield, Lander, Langston, Larson, Larson, LaVerghetta, Law, Lawrence, Lee, Lee, Lee, Leisenring, Leveille, Levenson, Levi, Levine, Lewis, Lewis, Lewis, Libralato, Lidon, Liebrecht, Lightsey, Lilly, Lim, Lim, Ling, Link, Link, Lipinski, Liu, Lo, Lobmeyer, Logue, Long, Long, Long, Long, L{\'{o}}pez-Caniego, Lotz, Love-Pruitt, Lubskiy, Luers, Luetgens, Luevano, {G. Flores Lui}, {Lund III}, Lundquist, Lunine, L{\"{u}}tzgendorf, Lynch, MacDonald, MacDonald, Macias, Macklis, Maghami, Maharaja, Maiolino, Makrygiannis, Malla, Malumuth, Manjavacas, Marini, Marrione,
  Marston, Martel, Martin, Martin, Martinez, Maschmann, Masci, Masetti, Maszkiewicz, Matthews, Matuskey, McBrayer, McCarthy, McCaughrean, McClare, McClare, McCloskey, McClurg, McCoy, McElwain, McGregor, McGuffey, McKay, McKenzie, McLean, McMaster, McNeil, {De Meester}, Mehalick, Meixner, Mel{\'{e}}ndez, Menzel, Menzel, Merz, Mesterharm, Meyer, Meyett, Meza, Midwinter, Milam, Miller, Miller, Miskey, Misselt, Mitchell, Mohan, Montoya, Moran, Morishita, Moro-Mart{\'{i}}n, Morrison, Morrison, Morse, Moschos, Moseley, Mosier, Mosner, Mountain, Muckenthaler, Mueller, Mueller, Muhiem, M{\"{u}}hlmann, Mullally, Mullen, Munger, Murphy, Murray, Muzerolle, Mycroft, Myers, Myers, {R. Myers}, Myers, Myrick, {Nagle, IV}, Nayak, Naylor, Neff, Nelan, Nella, Nguyen, Nguyen, Nickson, Nidhiry, Niedner, Nieto-Santisteban, Nikolov, Nishisaka, Noriega-Crespo, Nota, O'Mara, Oboryshko, O'Brien, Ochs, Offenberg, Ogle, Ohl, Olmsted, Osborne, O'Shaughnessy, {\"{O}}stlin, O'Sullivan, Otor, Ottens, Ouellette, Outlaw, Owens, Pacifici,
  Page, Paranilam, Park, Parrish, Paschal, Patapis, Patel, Patrick, {Pattishall Jr}, Paul, Paul, Pauly, Pavlovsky, Pe{\~{n}}a-Guerrero, Pedder, Peek, Pelham, Penanen, Perriello, Perrin, Perrine, Perrygo, Peslier, Petach, Peterson, Pfarr, Pierson, Pietraszkiewicz, Pilchen, Pipher, Pirzkal, Pitman, Player, Plesha, Plitzke, Pohner, Poletis, Pollizzi, Polster, Pontius, Pontoppidan, Porges, Potter, Prescott, Proffitt, Pueyo, {Quispe Neira}, Radich, Rager, Rameau, Ramey, Alarcon, Rampini, Rapp, Rashford, Rauscher, Ravindranath, Rawle, Rawlings, Ray, Regan, Rehm, Rehm, Reid, Reis, Renk, Reoch, Ressler, Rest, Reynolds, Richon, Richon, Ridgaway, Riedel, Rieke, Rieke, Rifelli, Rigby, Riggs, Ringel, Ritchie, Rix, Robberto, Robinson, Robinson, Robinson, Rock, Rodriguez, del Pino, Roellig, Rohrbach, Roman, Romelfanger, {Romo Jr}, Rosales, Rose, Roteliuk, Roth, Rothwell, Rouzaud, Rowe, Rowlands, Roy, Royer, Rui, Rumler, Rumpl, Russ, Ryan, Ryan, Saad, Sabata, Sabatino, Sabbi, Sabelhaus, Sabia, Sahu, Saif, Salvignol,
  Samara-Ratna, Samuelson, Sanders, Sappington, Sargent, Sauer, Savadkin, Sawicki, Schappell, Scheffer, Scheithauer, Scherer, Schiff, Schlawin, Schmeitzky, Schmitz, Schmude, Schneider, Schreiber, Schroeven-Deceuninck, Schultz, Schwab, Schwartz, Scoccimarro, Scott, Scott, Seaton, Seely, Seery, Seidleck, Sembach, Shanahan, Shaughnessy, Shaw, Shay, Sheehan, Sheth, Shih, Shivaei, Siegel, Sienkiewicz, Simmons, Simon, Sirianni, Sivaramakrishnan, Slade, Sloan, Slocum, Slowinski, Smith, Smith, Smith, Smith, Smith, Smith, Smolik, Soderblom, Sohn, Sokol, Sonneborn, Sontag, Sooy, Soummer, Southwood, Spain, Sparmo, Speer, Spencer, Sprofera, Stallcup, Stanley, Stansberry, Stark, Starr, Stassi, Steck, Steeley, Stephens, Stephenson, Stewart, Stiavelli, Jr, Strada, Straughn, Streetman, Strickland, Strobele, Stuhlinger, Stys, Such, Sukhatme, Sullivan, Sullivan, Sumner, Sun, Sunnquist, Swade, Swam, Swenton, Swoish, {Tam Litten}, Tamas, Tao, Taylor, Taylor, te~Plate, {Van Tea}, Teague, Telfer, Temim, Texter, Thatte, Thompson,
  Thompson, Thomson, Thronson, Tierney, Tikkanen, Tinnin, Tippet, Todd, Tran, Trauger, Trejo, {Vinh Truong}, Tsukamoto, Tufail, Tumlinson, Tustain, Tyra, Ubeda, Underwood, Uzzo, Vaclavik, Valenduc, Valenti, {Van Campen}, van~de Wetering, {Van Der Marel}, van Haarlem, Vandenbussche, van Dishoeck, Vanterpool, Vernoy, {Vila Costas}, Volk, Voorzaat, Voyton, Vydra, Waddy, Waelkens, Wahlgren, {Walker Jr}, Wander, Warfield, Warner, Wasiak, Wasiak, Wehner, Weiler, Weilert, Weiss, Wells, Welty, Wheate, Wheeler, White, Whitehouse, Whiteleather, Whitman, Williams, Willmer, Willott, Willoughby, Wilson, Wilson, Wilson, Windhorst, Wislowski, Wolfe, Wolfe, Wolff, Wondel, Woo, Woods, Worden, Workman, Wright, Wu, Wu, Wun, Wymer, Yadetie, Yan, Yang, Yates, Yeager, Yerger, Young, Young, Yu, Yu, Zak, Zeidler, Zepp, Zhou, Zincke, Zonak, \& Zondag}]{JWST2023}
Gardner, J.~P., Mather, J.~C., Abbott, R., {et~al.} 2023, Publications of the Astronomical Society of the Pacific, 135, 068001, \dodoi{10.1088/1538-3873/acd1b5}

\bibitem[{Glazebrook {et~al.}(2017)Glazebrook, Schreiber, Labb{\'{e}}, Nanayakkara, Kacprzak, Oesch, Papovich, Spitler, Straatman, Tran, \& Yuan}]{Glazebrook2017}
Glazebrook, K., Schreiber, C., Labb{\'{e}}, I., {et~al.} 2017, Nature, 544, 71, \dodoi{10.1038/nature21680}

\bibitem[{Greene {et~al.}(2006)Greene, Ho, \& Ulvestad}]{Greene2006}
Greene, J.~E., Ho, L.~C., \& Ulvestad, J.~S. 2006, The Astrophysical Journal, 636, 56, \dodoi{10.1086/497905}

\bibitem[{Haardt \& Madau(2012)}]{Haardt2012}
Haardt, F., \& Madau, P. 2012, The Astrophysical Journal, 746, 125, \dodoi{10.1088/0004-637X/746/2/125}

\bibitem[{Hahn \& Abel(2011)}]{Hahn2011}
Hahn, O., \& Abel, T. 2011, Monthly Notices of the Royal Astronomical Society, 415, 2101, \dodoi{10.1111/j.1365-2966.2011.18820.x}

\bibitem[{Harris {et~al.}(2020)Harris, Millman, van~der Walt, Gommers, Virtanen, Cournapeau, Wieser, Taylor, Berg, Smith, Kern, Picus, Hoyer, van Kerkwijk, Brett, Haldane, del R{\'{i}}o, Wiebe, Peterson, G{\'{e}}rard-Marchant, Sheppard, Reddy, Weckesser, Abbasi, Gohlke, \& Oliphant}]{Harris2020}
Harris, C.~R., Millman, K.~J., van~der Walt, S.~J., {et~al.} 2020, Nature, 585, 357, \dodoi{10.1038/s41586-020-2649-2}

\bibitem[{Hashimoto {et~al.}(2018)Hashimoto, Laporte, Mawatari, Ellis, Inoue, Zackrisson, Roberts-Borsani, Zheng, Tamura, Bauer, Fletcher, Harikane, Hatsukade, Hayatsu, Matsuda, Matsuo, Okamoto, Ouchi, Pell{\'{o}}, Rydberg, Shimizu, Taniguchi, Umehata, \& Yoshida}]{Hasimoto2018}
Hashimoto, T., Laporte, N., Mawatari, K., {et~al.} 2018, Nature, 557, 392, \dodoi{10.1038/s41586-018-0117-z}

\bibitem[{Henden {et~al.}(2018)Henden, Puchwein, Shen, \& Sijacki}]{Henden2018}
Henden, N.~A., Puchwein, E., Shen, S., \& Sijacki, D. 2018, Monthly Notices of the Royal Astronomical Society, 479, 5385, \dodoi{10.1093/mnras/sty1780}

\bibitem[{Higuchi {et~al.}(2019)Higuchi, Ouchi, Ono, Shibuya, Toshikawa, Harikane, Kojima, Chiang, Egami, Kashikawa, Overzier, Konno, Inoue, Hasegawa, Fujimoto, Goto, Ishikawa, Ito, Komiyama, \& Tanaka}]{Higuchi2019}
Higuchi, R., Ouchi, M., Ono, Y., {et~al.} 2019, The Astrophysical Journal, 879, 28, \dodoi{10.3847/1538-4357/ab2192}

\bibitem[{Hill {et~al.}(2022)Hill, Chapman, Phadke, Aravena, Archipley, Ashby, B{\'{e}}thermin, Canning, Gonzalez, Greve, Gururajan, Hayward, Hezaveh, Jarugula, MacIntyre, Marrone, Miller, Rennehan, Reuter, Rotermund, Scott, Spilker, Vieira, Wang, \& Wei{\ss}}]{Hill2022}
Hill, R., Chapman, S., Phadke, K.~A., {et~al.} 2022, Monthly Notices of the Royal Astronomical Society, 512, 4352, \dodoi{10.1093/mnras/stab3539}

\bibitem[{Hopkins(2015)}]{Hopkins2015a}
Hopkins, P.~F. 2015, Monthly Notices of the Royal Astronomical Society, 450, 53, \dodoi{10.1093/mnras/stv195}

\bibitem[{Hopkins(2017)}]{Hopkins2016b}
---. 2017, Monthly Notices of the Royal Astronomical Society, 466, 3387, \dodoi{10.1093/mnras/stw3306}

\bibitem[{Hopkins {et~al.}(2014)Hopkins, Kere{\v{s}}, O{\~{n}}orbe, Faucher-Gigu{\`{e}}re, Quataert, Murray, \& Bullock}]{Hopkins2014}
Hopkins, P.~F., Kere{\v{s}}, D., O{\~{n}}orbe, J., {et~al.} 2014, Monthly Notices of the Royal Astronomical Society, 445, 581, \dodoi{10.1093/mnras/stu1738}

\bibitem[{Hopkins {et~al.}(2010)Hopkins, Murray, Quataert, \& Thompson}]{Hopkins2010b}
Hopkins, P.~F., Murray, N., Quataert, E., \& Thompson, T.~A. 2010, Monthly Notices of the Royal Astronomical Society: Letters, 401, L19, \dodoi{10.1111/j.1745-3933.2009.00777.x}

\bibitem[{Hopkins \& Quataert(2011)}]{Hopkins2011}
Hopkins, P.~F., \& Quataert, E. 2011, Monthly Notices of the Royal Astronomical Society, 415, 1027, \dodoi{10.1111/j.1365-2966.2011.18542.x}

\bibitem[{Hopkins {et~al.}(2017)Hopkins, Wetzel, Keres, Faucher-Giguere, Quataert, Boylan-Kolchin, Murray, Hayward, Garrison-Kimmel, Hummels, Feldmann, Torrey, Ma, Angles-Alcazar, Su, Orr, Schmitz, Escala, Sanderson, Grudic, Hafen, Kim, Fitts, Bullock, Wheeler, Chan, Elbert, \& Narananan}]{Hopkins2017}
Hopkins, P.~F., Wetzel, A., Keres, D., {et~al.} 2017, 000, \dodoi{10.1093/mnras/sty1690}

\bibitem[{Hough {et~al.}(2023)Hough, Rennehan, Kobayashi, Loubser, Dav{\'{e}}, Babul, \& Cui}]{Hough2023}
Hough, R.~T., Rennehan, D., Kobayashi, C., {et~al.} 2023, Monthly Notices of the Royal Astronomical Society, 525, 1061, \dodoi{10.1093/mnras/stad2394}

\bibitem[{Hunter(2007)}]{Hunter2007}
Hunter, J.~D. 2007, Computing in Science and Engineering, 9, 90, \dodoi{10.1109/MCSE.2007.55}

\bibitem[{Ishigaki {et~al.}(2016)Ishigaki, Ouchi, \& Harikane}]{Ishigaki2015}
Ishigaki, M., Ouchi, M., \& Harikane, Y. 2016, The Astrophysical Journal, 822, 5, \dodoi{10.3847/0004-637X/822/1/5}

\bibitem[{Ito {et~al.}(2019)Ito, Kashikawa, Toshikawa, Overzier, Tanaka, Kubo, Shibuya, Ishikawa, Onoue, Uchiyama, Liang, Higuchi, Martin, Lee, Komiyama, \& Huang}]{Ito2019}
Ito, K., Kashikawa, N., Toshikawa, J., {et~al.} 2019, The Astrophysical Journal, 878, 68, \dodoi{10.3847/1538-4357/ab1f0c}

\bibitem[{Iwamoto {et~al.}(1999)Iwamoto, Brachwitz, Nomoto, Kishimoto, Umeda, Hix, \& Thielemann}]{Iwamoto1999}
Iwamoto, K., Brachwitz, F., Nomoto, K., {et~al.} 1999, The Astrophysical Journal Supplement Series, 125, 439, \dodoi{10.1086/313278}

\bibitem[{Jiang {et~al.}(2018)Jiang, Wu, Bian, Chiang, Ho, Shen, Zheng, Bailey, Blanc, Crane, Fan, Mateo, Olszewski, Oyarz{\'{u}}n, Wang, \& Wu}]{Jiang2018}
Jiang, L., Wu, J., Bian, F., {et~al.} 2018, Nature Astronomy, 2, 962, \dodoi{10.1038/s41550-018-0587-9}

\bibitem[{Jin {et~al.}(2024)Jin, Sillassen, Magdis, Brinch, Shuntov, Brammer, Gobat, Valentino, Carnall, Lee, Vijayan, Gillman, Kokorev, {Le Bail}, Greve, Gullberg, Gould, \& Toft}]{Jin2024}
Jin, S., Sillassen, N.~B., Magdis, G.~E., {et~al.} 2024, Astronomy {\&} Astrophysics, 683, L4, \dodoi{10.1051/0004-6361/202348540}

\bibitem[{Joung {et~al.}(2009)Joung, Cen, \& Bryan}]{RyanJoung2009}
Joung, M.~R., Cen, R., \& Bryan, G.~L. 2009, The Astrophysical Journal, 692, L1, \dodoi{10.1088/0004-637X/692/1/L1}

\bibitem[{Kakimoto {et~al.}(2024)Kakimoto, Tanaka, Onodera, Shimakawa, Wu, Gould, Ito, Jin, Kubo, Suzuki, Toft, Valentino, \& Yabe}]{Kakimoto2024}
Kakimoto, T., Tanaka, M., Onodera, M., {et~al.} 2024, The Astrophysical Journal, 963, 49, \dodoi{10.3847/1538-4357/ad1ff1}

\bibitem[{Katz \& White(1993)}]{Katz1993}
Katz, N., \& White, S. D.~M. 1993, The Astrophysical Journal, 412, 455, \dodoi{10.1086/172935}

\bibitem[{Keller {et~al.}(2023)Keller, Munshi, Trebitsch, \& Tremmel}]{Keller2023}
Keller, B.~W., Munshi, F., Trebitsch, M., \& Tremmel, M. 2023, The Astrophysical Journal Letters, 943, L28, \dodoi{10.3847/2041-8213/acb148}

\bibitem[{Kimmig {et~al.}(2023)Kimmig, Remus, Seidel, Valenzuela, Dolag, \& Burkert}]{Kimmig2024}
Kimmig, L.~C., Remus, R.-S., Seidel, B., {et~al.} 2023, arXiv preprint arXiv:2310.16085.
\newblock \doarXiv{2310.16085}

\bibitem[{King \& Pounds(2015)}]{King2015}
King, A., \& Pounds, K. 2015, Annual Review of Astronomy and Astrophysics, 53, 115, \dodoi{10.1146/annurev-astro-082214-122316}

\bibitem[{Kocevski {et~al.}(2023)Kocevski, Onoue, Inayoshi, Trump, {Arrabal Haro}, Grazian, Dickinson, Finkelstein, Kartaltepe, Hirschmann, Aird, Holwerda, Fujimoto, Juneau, Amor{\'{i}}n, Backhaus, Bagley, Barro, Bell, Bisigello, Calabr{\`{o}}, Cleri, Cooper, Ding, Grogin, Ho, Hutchison, Inoue, Jiang, Jones, Koekemoer, Li, Li, McGrath, Molina, Papovich, P{\'{e}}rez-Gonz{\'{a}}lez, Pirzkal, Wilkins, Yang, \& Yung}]{Kocevski2023}
Kocevski, D.~D., Onoue, M., Inayoshi, K., {et~al.} 2023, The Astrophysical Journal Letters, 954, L4, \dodoi{10.3847/2041-8213/ace5a0}

\bibitem[{Kormendy \& Ho(2013)}]{Kormendy2013}
Kormendy, J., \& Ho, L.~C. 2013, Annual Review of Astronomy and Astrophysics, 51, 511, \dodoi{10.1146/annurev-astro-082708-101811}

\bibitem[{Kravtsov \& Borgani(2012)}]{Kravtsov2012}
Kravtsov, A.~V., \& Borgani, S. 2012, Annual Review of Astronomy and Astrophysics, 50, 353, \dodoi{10.1146/annurev-astro-081811-125502}

\bibitem[{Krumholz {et~al.}(2009)Krumholz, McKee, \& Tumlinson}]{Krumholz2009}
Krumholz, M.~R., McKee, C.~F., \& Tumlinson, J. 2009, The Astrophysical Journal, 699, 850, \dodoi{10.1088/0004-637X/699/1/850}

\bibitem[{Labb{\'{e}} {et~al.}(2023)Labb{\'{e}}, van Dokkum, Nelson, Bezanson, Suess, Leja, Brammer, Whitaker, Mathews, Stefanon, \& Wang}]{Labbe2023}
Labb{\'{e}}, I., van Dokkum, P., Nelson, E., {et~al.} 2023, Nature, 616, 266, \dodoi{10.1038/s41586-023-05786-2}

\bibitem[{Lambrides {et~al.}(2024)Lambrides, Chiaberge, Long, Liu, Akins, Ptak, Andika, Capetti, Casey, Champagne, Chworowsky, Clarke, Cooper, Ding, Dong, Faisst, Forman, Franco, Gillman, Gozaliasl, Hall, Harish, Hayward, Hirschmann, Hutchison, Jahnke, Jin, Kartaltepe, Kleiner, Koekemoer, Kokorev, Manning, Martin, McKinney, Norman, Nyland, Onoue, Robertson, Shuntov, Silverman, Stiavelli, Trakhtenbrot, Vardoulaki, Zavala, Allen, Ilbert, McCracken, Paquereau, Rhodes, \& Toft}]{Lambrides2024}
Lambrides, E., Chiaberge, M., Long, A.~S., {et~al.} 2024, The Astrophysical Journal Letters, 961, L25, \dodoi{10.3847/2041-8213/ad11ee}

\bibitem[{Lammers {et~al.}(2023)Lammers, Iyer, Ibarra-Medel, Pacifici, S{\'{a}}nchez, Tacchella, \& Woo}]{Lammers2023}
Lammers, C., Iyer, K.~G., Ibarra-Medel, H., {et~al.} 2023, The Astrophysical Journal, 953, 26, \dodoi{10.3847/1538-4357/acdd57}

\bibitem[{Lanson \& Vila(2008{\natexlab{a}})}]{Lanson2008a}
Lanson, N., \& Vila, J.-P. 2008{\natexlab{a}}, SIAM Journal on Numerical Analysis, 46, 1912, \dodoi{10.1137/S0036142903427718}

\bibitem[{Lanson \& Vila(2008{\natexlab{b}})}]{Lanson2008b}
---. 2008{\natexlab{b}}, SIAM Journal on Numerical Analysis, 46, 1935, \dodoi{10.1137/S003614290444739X}

\bibitem[{Laor \& Netzer(1989)}]{Laor1989}
Laor, A., \& Netzer, H. 1989, Monthly Notices of the Royal Astronomical Society, 238, 897, \dodoi{10.1093/mnras/238.3.897}

\bibitem[{Lim {et~al.}(2021)Lim, Scott, Babul, Barnes, Kay, Mccarthy, Rennehan, \& Vogelsberger}]{Lim2020}
Lim, S., Scott, D., Babul, A., {et~al.} 2021, Monthly Notices of the Royal Astronomical Society, 501, 1803, \dodoi{10.1093/mnras/staa3693}

\bibitem[{Lim {et~al.}(2024)Lim, Tacchella, Schaye, Schaller, Helton, Kugel, \& Maiolino}]{Lim2024}
Lim, S., Tacchella, S., Schaye, J., {et~al.} 2024, arXiv preprint arXiv:2402.17819.
\newblock \doarXiv{2402.17819}

\bibitem[{Lovell {et~al.}(2021)Lovell, Geach, Dav{\'{e}}, Narayanan, \& Li}]{Lovell2021}
Lovell, C.~C., Geach, J.~E., Dav{\'{e}}, R., Narayanan, D., \& Li, Q. 2021, Monthly Notices of the Royal Astronomical Society, 502, 772, \dodoi{10.1093/mnras/staa4043}

\bibitem[{Lovell {et~al.}(2020)Lovell, Vijayan, Thomas, Wilkins, Barnes, Irodotou, \& Roper}]{Lovell2020}
Lovell, C.~C., Vijayan, A.~P., Thomas, P.~A., {et~al.} 2020, Monthly Notices of the Royal Astronomical Society, 500, 2127, \dodoi{10.1093/mnras/staa3360}

\bibitem[{Maccarone {et~al.}(2003)Maccarone, Gallo, \& Fender}]{Maccarone2003}
Maccarone, T.~J., Gallo, E., \& Fender, R. 2003, Monthly Notices of the Royal Astronomical Society, 345, L19, \dodoi{10.1046/j.1365-8711.2003.07161.x}

\bibitem[{Madau \& Dickinson(2014)}]{Madau2014}
Madau, P., \& Dickinson, M. 2014, Annual Review of Astronomy and Astrophysics, 52, 415, \dodoi{10.1146/annurev-astro-081811-125615}

\bibitem[{Madau {et~al.}(2014)Madau, Haardt, \& Dotti}]{Madau2014a}
Madau, P., Haardt, F., \& Dotti, M. 2014, The Astrophysical Journal, 784, L38, \dodoi{10.1088/2041-8205/784/2/L38}

\bibitem[{Mart{\'{i}}nez-Aldama {et~al.}(2018)Mart{\'{i}}nez-Aldama, del Olmo, Marziani, Sulentic, Negrete, Dultzin, D'Onofrio, \& Perea}]{MartinezAldama2018}
Mart{\'{i}}nez-Aldama, M.~L., del Olmo, A., Marziani, P., {et~al.} 2018, Astronomy {\&} Astrophysics, 618, A179, \dodoi{10.1051/0004-6361/201833541}

\bibitem[{McClintock {et~al.}(2006)McClintock, Shafee, Narayan, Remillard, Davis, \& Li}]{McClintock2006}
McClintock, J.~E., Shafee, R., Narayan, R., {et~al.} 2006, The Astrophysical Journal, 652, 518, \dodoi{10.1086/508457}

\bibitem[{Merlin {et~al.}(2019)Merlin, Fortuni, Torelli, Santini, Castellano, Fontana, Grazian, Pentericci, Pilo, \& Schmidt}]{Merlin2019}
Merlin, E., Fortuni, F., Torelli, M., {et~al.} 2019, Monthly Notices of the Royal Astronomical Society, 490, 3309, \dodoi{10.1093/mnras/stz2615}

\bibitem[{Miller {et~al.}(2018)Miller, Chapman, Aravena, Ashby, Hayward, Vieira, Wei{\ss}, Babul, B{\'{e}}thermin, Bradford, Brodwin, Carlstrom, Chen, Cunningham, {De Breuck}, Gonzalez, Greve, Harnett, Hezaveh, Lacaille, Litke, Ma, Malkan, Marrone, Morningstar, Murphy, Narayanan, Pass, Perry, Phadke, Rennehan, Rotermund, Simpson, Spilker, Sreevani, Stark, Strandet, \& Strom}]{Miller2018}
Miller, T.~B., Chapman, S.~C., Aravena, M., {et~al.} 2018, Nature, 556, 469, \dodoi{10.1038/s41586-018-0025-2}

\bibitem[{Momcheva {et~al.}(2016)Momcheva, Brammer, van Dokkum, Skelton, Whitaker, Nelson, Fumagalli, Maseda, Leja, Franx, Rix, Bezanson, Cunha, Dickey, Schreiber, Illingworth, Kriek, Labb{\'{e}}, Lange, Lundgren, Magee, Marchesini, Oesch, Pacifici, Patel, Price, Tal, Wake, van~der Wel, \& Wuyts}]{Momcheva2016}
Momcheva, I.~G., Brammer, G.~B., van Dokkum, P.~G., {et~al.} 2016, The Astrophysical Journal Supplement Series, 225, 27, \dodoi{10.3847/0067-0049/225/2/27}

\bibitem[{Moster {et~al.}(2011)Moster, Somerville, Newman, \& Rix}]{Moster2011}
Moster, B.~P., Somerville, R.~S., Newman, J.~A., \& Rix, H.-W. 2011, The Astrophysical Journal, 731, 113, \dodoi{10.1088/0004-637X/731/2/113}

\bibitem[{Muratov {et~al.}(2015)Muratov, Kere{\v{s}}, Faucher-Gigu{\`{e}}re, Hopkins, Quataert, \& Murray}]{Muratov2015}
Muratov, A.~L., Kere{\v{s}}, D., Faucher-Gigu{\`{e}}re, C.-A., {et~al.} 2015, Monthly Notices of the Royal Astronomical Society, 454, 2691, \dodoi{10.1093/mnras/stv2126}

\bibitem[{Muzzin {et~al.}(2009)Muzzin, Wilson, Yee, Hoekstra, Gilbank, Surace, Lacy, Blindert, Majumdar, Demarco, Gardner, Gladders, \& Lonsdale}]{Muzzin2009}
Muzzin, A., Wilson, G., Yee, H. K.~C., {et~al.} 2009, The Astrophysical Journal, 698, 1934, \dodoi{10.1088/0004-637X/698/2/1934}

\bibitem[{Nelson {et~al.}(2024)Nelson, Pillepich, Ayromlou, Lee, Lehle, Rohr, \& Truong}]{Nelson2024}
Nelson, D., Pillepich, A., Ayromlou, M., {et~al.} 2024, arXiv preprint arXiv:2311.06338.
\newblock \doarXiv{2311.06338}

\bibitem[{Nomoto {et~al.}(2006)Nomoto, Tominaga, Umeda, Kobayashi, \& Maeda}]{Nomoto2006}
Nomoto, K., Tominaga, N., Umeda, H., Kobayashi, C., \& Maeda, K. 2006, Nuclear Physics A, 777, 424, \dodoi{10.1016/j.nuclphysa.2006.05.008}

\bibitem[{Noordeh {et~al.}(2021)Noordeh, Canning, Willis, Allen, Mantz, Stanford, \& Brammer}]{Noordeh2021}
Noordeh, E., Canning, R. E.~A., Willis, J.~P., {et~al.} 2021, Monthly Notices of the Royal Astronomical Society, 507, 5272, \dodoi{10.1093/mnras/stab2459}

\bibitem[{Oesch {et~al.}(2015)Oesch, {Van Dokkum}, Illingworth, Bouwens, Momcheva, Holden, Roberts-Borsani, Smit, Franx, Labb{\'{e}}, Gonz{\'{a}}lez, \& Magee}]{Oesch2015}
Oesch, P.~A., {Van Dokkum}, P.~G., Illingworth, G.~D., {et~al.} 2015, Astrophysical Journal Letters, 804, \dodoi{10.1088/2041-8205/804/2/L30}

\bibitem[{Oesch {et~al.}(2023)Oesch, Brammer, Naidu, Bouwens, Chisholm, Illingworth, Matthee, Nelson, Qin, Reddy, Shapley, Shivaei, van Dokkum, Weibel, Whitaker, Wuyts, Covelo-Paz, Endsley, Fudamoto, Giovinazzo, Herard-Demanche, Kerutt, Kramarenko, Labbe, Leonova, Lin, Magee, Marchesini, Maseda, Mason, Matharu, Meyer, Neufeld, {Prieto Lyon}, Schaerer, Sharma, Shuntov, Smit, Stefanon, Wyithe, \& Xiao}]{Oesch2023}
Oesch, P.~A., Brammer, G., Naidu, R.~P., {et~al.} 2023, Monthly Notices of the Royal Astronomical Society, 525, 2864, \dodoi{10.1093/mnras/stad2411}

\bibitem[{O{\~{n}}orbe {et~al.}(2014)O{\~{n}}orbe, Garrison-Kimmel, Maller, Bullock, Rocha, \& Hahn}]{Onorbe2013}
O{\~{n}}orbe, J., Garrison-Kimmel, S., Maller, A.~H., {et~al.} 2014, Monthly Notices of the Royal Astronomical Society, 437, 1894, \dodoi{10.1093/mnras/stt2020}

\bibitem[{Oppenheimer \& Dav{\'{e}}(2008)}]{Oppenheimer2008}
Oppenheimer, B.~D., \& Dav{\'{e}}, R. 2008, Monthly Notices of the Royal Astronomical Society, 387, 577, \dodoi{10.1111/j.1365-2966.2008.13280.x}

\bibitem[{Pacifici {et~al.}(2016)Pacifici, Kassin, Weiner, Holden, Gardner, Faber, Ferguson, Koo, Primack, Bell, Dekel, Gawiser, Giavalisco, Rafelski, Simons, Barro, Croton, Dav{\'{e}}, Fontana, Grogin, Koekemoer, Lee, Salmon, Somerville, \& Behroozi}]{Pacifici2016}
Pacifici, C., Kassin, S.~A., Weiner, B.~J., {et~al.} 2016, The Astrophysical Journal, 832, 79, \dodoi{10.3847/0004-637X/832/1/79}

\bibitem[{Pakmor {et~al.}(2023)Pakmor, Springel, Coles, Guillet, Pfrommer, Bose, Barrera, Delgado, Ferlito, Frenk, Hadzhiyska, Hern{\'{a}}ndez-Aguayo, Hernquist, Kannan, \& White}]{Pakmor2023}
Pakmor, R., Springel, V., Coles, J.~P., {et~al.} 2023, Monthly Notices of the Royal Astronomical Society, 524, 2539, \dodoi{10.1093/mnras/stac3620}

\bibitem[{Parsotan {et~al.}(2020)Parsotan, Cochrane, Hayward, Angl{\'{e}}s-Alc{\'{a}}zar, Feldmann, Faucher-Gigu{\`{e}}re, Wellons, \& Hopkins}]{Parsotan2020}
Parsotan, T., Cochrane, R.~K., Hayward, C.~C., {et~al.} 2020, Monthly Notices of the Royal Astronomical Society, 501, 1591, \dodoi{10.1093/mnras/staa3765}

\bibitem[{Perez \& Granger(2007)}]{Perez2007}
Perez, F., \& Granger, B.~E. 2007, Computing in Science and Engineering, 9, 21, \dodoi{10.1109/MCSE.2007.53}

\bibitem[{Perna {et~al.}(2017)Perna, Lanzuisi, Brusa, Cresci, \& Mignoli}]{Perna2017a}
Perna, M., Lanzuisi, G., Brusa, M., Cresci, G., \& Mignoli, M. 2017, Astronomy {\&} Astrophysics, 606, A96, \dodoi{10.1051/0004-6361/201730819}

\bibitem[{{Planck Collaboration XIII}(2015)}]{Ade2016}
{Planck Collaboration XIII}. 2015, Astronomy {\&} Astrophysics, 594, A13, \dodoi{10.1051/0004-6361/201525830}

\bibitem[{Price \& Monaghan(2007)}]{Price2007}
Price, D.~J., \& Monaghan, J.~J. 2007, Monthly Notices of the Royal Astronomical Society, 374, 1347, \dodoi{10.1111/j.1365-2966.2006.11241.x}

\bibitem[{Remus {et~al.}(2023)Remus, Dolag, \& Dannerbauer}]{Remus2023}
Remus, R.-S., Dolag, K., \& Dannerbauer, H. 2023, The Astrophysical Journal, 950, 191, \dodoi{10.3847/1538-4357/accb91}

\bibitem[{Rennehan {et~al.}(2020)Rennehan, Babul, Hayward, Bottrell, Hani, \& Chapman}]{Rennehan2020}
Rennehan, D., Babul, A., Hayward, C.~C., {et~al.} 2020, Monthly Notices of the Royal Astronomical Society, 493, 4607, \dodoi{10.1093/mnras/staa541}

\bibitem[{Rieke {et~al.}(2023)Rieke, Kelly, Misselt, Stansberry, Boyer, Beatty, Egami, Florian, Greene, Hainline, Leisenring, Roellig, Schlawin, Sun, Tinnin, Williams, Willmer, Wilson, Clark, Rohrbach, Brooks, Canipe, Correnti, DiFelice, Gennaro, Girard, Hartig, Hilbert, Koekemoer, Nikolov, Pirzkal, Rest, Robberto, Sunnquist, Telfer, Wu, Ferry, Lewis, Baum, Beichman, Doyon, Dressler, Eisenstein, Ferrarese, Hodapp, Horner, Jaffe, Johnstone, Krist, Martin, McCarthy, Meyer, Rieke, Trauger, \& Young}]{NIRCam2023}
Rieke, M.~J., Kelly, D.~M., Misselt, K., {et~al.} 2023, Publications of the Astronomical Society of the Pacific, 135, 028001, \dodoi{10.1088/1538-3873/acac53}

\bibitem[{Robertson {et~al.}(2023)Robertson, Tacchella, Johnson, Hainline, Whitler, Eisenstein, Endsley, Rieke, Stark, Alberts, Dressler, Egami, Hausen, Rieke, Shivaei, Williams, Willmer, Arribas, Bonaventura, Bunker, Cameron, Carniani, Charlot, Chevallard, Curti, Curtis-Lake, D'Eugenio, Jakobsen, Looser, L{\"{u}}tzgendorf, Maiolino, Maseda, Rawle, Rix, Smit, {\"{U}}bler, Willott, Witstok, Baum, Bhatawdekar, Boyett, Chen, de~Graaff, Florian, Helton, Hviding, Ji, Kumari, Lyu, Nelson, Sandles, Saxena, Suess, Sun, Topping, \& Wallace}]{Robertson2023}
Robertson, B.~E., Tacchella, S., Johnson, B.~D., {et~al.} 2023, Nature Astronomy, 7, 611, \dodoi{10.1038/s41550-023-01921-1}

\bibitem[{S{\c{a}}dowski(2009)}]{Sadowski2009}
S{\c{a}}dowski, A. 2009, The Astrophysical Journal Supplement Series, 183, 171, \dodoi{10.1088/0067-0049/183/2/171}

\bibitem[{Sazonov {et~al.}(2005)Sazonov, Ostriker, Ciotti, \& Sunyaev}]{Sazonov2005}
Sazonov, S.~Y., Ostriker, J.~P., Ciotti, L., \& Sunyaev, R.~A. 2005, Monthly Notices of the Royal Astronomical Society, 358, 168, \dodoi{10.1111/J.1365-2966.2005.08763.x}

\bibitem[{Scannapieco \& Bildsten(2005)}]{Scannapieco2005}
Scannapieco, E., \& Bildsten, L. 2005, The Astrophysical Journal, 629, L85, \dodoi{10.1086/452632}

\bibitem[{Scharr{\'{e}} {et~al.}(2024)Scharr{\'{e}}, Sorini, \& Dav{\'{e}}}]{Sharre2024}
Scharr{\'{e}}, L., Sorini, D., \& Dav{\'{e}}, R. 2024, arXiv preprint arXiv:2404.07252.
\newblock \doarXiv{2404.07252}

\bibitem[{Schaye {et~al.}(2023)Schaye, Kugel, Schaller, Helly, Braspenning, Elbers, McCarthy, van Daalen, Vandenbroucke, Frenk, Kwan, Salcido, Bah{\'{e}}, Borrow, Chaikin, Hahn, Hu{\v{s}}ko, Jenkins, Lacey, \& Nobels}]{Schaye2023}
Schaye, J., Kugel, R., Schaller, M., {et~al.} 2023, Monthly Notices of the Royal Astronomical Society, 526, 4978, \dodoi{10.1093/mnras/stad2419}

\bibitem[{Setton {et~al.}(2024)Setton, Khullar, Miller, Bezanson, Greene, Suess, Whitaker, Antwi-Danso, Atek, Brammer, Cutler, Dayal, Feldmann, Furtak, Fujimoto, Glazebrook, Goulding, Kokorev, Labbe, Leja, Ma, Marchesini, Nanayakkara, Pan, Price, Siegel, Shipley, Weaver, van Dokkum, Wang, \& Williams}]{Setton2024}
Setton, D.~J., Khullar, G., Miller, T.~B., {et~al.} 2024, arXiv preprint arXiv:2402.05664.
\newblock \doarXiv{2402.05664}

\bibitem[{Smith {et~al.}(2017)Smith, Bryan, Glover, Goldbaum, Turk, Regan, Wise, Schive, Abel, Emerick, O'Shea, Anninos, Hummels, \& Khochfar}]{Smith2017}
Smith, B.~D., Bryan, G.~L., Glover, S. C.~O., {et~al.} 2017, Monthly Notices of the Royal Astronomical Society, 466, 2217, \dodoi{10.1093/mnras/stw3291}

\bibitem[{Somerville {et~al.}(2004)Somerville, Lee, Ferguson, Gardner, Moustakas, \& Giavalisco}]{Somerville2004}
Somerville, R.~S., Lee, K., Ferguson, H.~C., {et~al.} 2004, The Astrophysical Journal, 600, L171, \dodoi{10.1086/378628}

\bibitem[{Springel {et~al.}(2021)Springel, Pakmor, Zier, \& Reinecke}]{Springel2021}
Springel, V., Pakmor, R., Zier, O., \& Reinecke, M. 2021, Monthly Notices of the Royal Astronomical Society, 506, 2871, \dodoi{10.1093/mnras/stab1855}

\bibitem[{Stott {et~al.}(2011)Stott, Collins, Burke, Hamilton-Morris, \& Smith}]{Stott2011}
Stott, J.~P., Collins, C.~A., Burke, C., Hamilton-Morris, V., \& Smith, G.~P. 2011, Monthly Notices of the Royal Astronomical Society, 414, 445, \dodoi{10.1111/j.1365-2966.2011.18404.x}

\bibitem[{Straatman {et~al.}(2015)Straatman, Labb{\'{e}}, Spitler, Glazebrook, Tomczak, Allen, Brammer, Cowley, Dokkum, Kacprzak, Kawinwanichakij, Mehrtens, Nanayakkara, Papovich, Persson, Quadri, Rees, Tilvi, Tran, \& Whitaker}]{Straatman2015}
Straatman, C. M.~S., Labb{\'{e}}, I., Spitler, L.~R., {et~al.} 2015, The Astrophysical Journal, 808, L29, \dodoi{10.1088/2041-8205/808/1/L29}

\bibitem[{Szpila {et~al.}(2024)Szpila, Dav{\'{e}}, Rennehan, Cui, \& Hough}]{Spzila2024}
Szpila, J., Dav{\'{e}}, R., Rennehan, D., Cui, W., \& Hough, R. 2024, arXiv preprint arXiv:2402.08729.
\newblock \doarXiv{2402.08729}

\bibitem[{Trebitsch {et~al.}(2021)Trebitsch, Dubois, Volonteri, Pfister, Cadiou, Katz, Rosdahl, Kimm, Pichon, Beckmann, Devriendt, \& Slyz}]{Trebitsch2021}
Trebitsch, M., Dubois, Y., Volonteri, M., {et~al.} 2021, Astronomy {\&} Astrophysics, 653, A154, \dodoi{10.1051/0004-6361/202037698}

\bibitem[{Turk {et~al.}(2011)Turk, Smith, Oishi, Skory, Skillman, Abel, \& Norman}]{Turk2011}
Turk, M.~J., Smith, B.~D., Oishi, J.~S., {et~al.} 2011, The Astrophysical Journal Supplement Series, 192, 9, \dodoi{10.1088/0067-0049/192/1/9}

\bibitem[{Valentino {et~al.}(2020)Valentino, Tanaka, Davidzon, Toft, G{\'{o}}mez-Guijarro, Stockmann, Onodera, Brammer, Ceverino, Faisst, Gallazzi, Hayward, Ilbert, Kubo, Magdis, Selsing, Shimakawa, Sparre, Steinhardt, Yabe, \& Zabl}]{Valentino2020}
Valentino, F., Tanaka, M., Davidzon, I., {et~al.} 2020, The Astrophysical Journal, 889, 93, \dodoi{10.3847/1538-4357/ab64dc}

\bibitem[{Valentino {et~al.}(2023)Valentino, Brammer, Gould, Kokorev, Fujimoto, Jespersen, Vijayan, Weaver, Ito, Tanaka, Ilbert, Magdis, Whitaker, Faisst, Gallazzi, Gillman, Gim{\'{e}}nez-Arteaga, G{\'{o}}mez-Guijarro, Kubo, Heintz, Hirschmann, Oesch, Onodera, Rizzo, Lee, Strait, \& Toft}]{Valentino2023}
Valentino, F., Brammer, G., Gould, K. M.~L., {et~al.} 2023, The Astrophysical Journal, 947, 20, \dodoi{10.3847/1538-4357/acbefa}

\bibitem[{Virtanen {et~al.}(2020)Virtanen, Gommers, Oliphant, Haberland, Reddy, Cournapeau, Burovski, Peterson, Weckesser, Bright, van~der Walt, Brett, Wilson, Millman, Mayorov, Nelson, Jones, Kern, Larson, Carey, Polat, Feng, Moore, VanderPlas, Laxalde, Perktold, Cimrman, Henriksen, Quintero, Harris, Archibald, Ribeiro, Pedregosa, \& van Mulbregt}]{Virtanen2020}
Virtanen, P., Gommers, R., Oliphant, T.~E., {et~al.} 2020, Nature Methods, 17, 261, \dodoi{10.1038/s41592-019-0686-2}

\bibitem[{Vogelsberger {et~al.}(2014)Vogelsberger, Genel, Springel, Torrey, Sijacki, Xu, Snyder, Nelson, \& Hernquist}]{Vogelsberger2014}
Vogelsberger, M., Genel, S., Springel, V., {et~al.} 2014, Monthly Notices of the Royal Astronomical Society, 444, 1518, \dodoi{10.1093/mnras/stu1536}

\bibitem[{Webb {et~al.}(2015)Webb, Muzzin, Noble, Bonaventura, Geach, Hezevah, Lidman, Wilson, Yee, Surace, \& Shupe}]{Webb2015}
Webb, T. M.~A., Muzzin, A., Noble, A., {et~al.} 2015, The Astrophysical Journal, 814, 96, \dodoi{10.1088/0004-637X/814/2/96}

\bibitem[{Whiley {et~al.}(2008)Whiley, Arag{\'{o}}n-Salamanca, {De Lucia}, von~der Linden, Bamford, Best, Bremer, Jablonka, Johnson, Milvang-Jensen, Noll, Poggianti, Rudnick, Saglia, White, \& Zaritsky}]{Whiley2008}
Whiley, I.~M., Arag{\'{o}}n-Salamanca, A., {De Lucia}, G., {et~al.} 2008, Monthly Notices of the Royal Astronomical Society, 387, 1253, \dodoi{10.1111/j.1365-2966.2008.13324.x}

\bibitem[{White \& Rees(1978)}]{White1978}
White, S. D.~M., \& Rees, M.~J. 1978, Monthly Notices of the Royal Astronomical Society, 183, 341, \dodoi{10.1093/mnras/183.3.341}

\bibitem[{Willis {et~al.}(2020)Willis, Canning, Noordeh, Allen, King, Mantz, Morris, Stanford, \& Brammer}]{Willis2020}
Willis, J.~P., Canning, R. E.~A., Noordeh, E.~S., {et~al.} 2020, Nature, 577, 39, \dodoi{10.1038/s41586-019-1829-4}

\bibitem[{Wilson {et~al.}(2009)Wilson, Muzzin, Yee, Lacy, Surace, Gilbank, Blindert, Hoekstra, Majumdar, Demarco, Gardner, Gladders, \& Lonsdale}]{Wilson2009}
Wilson, G., Muzzin, A., Yee, H. K.~C., {et~al.} 2009, The Astrophysical Journal, 698, 1943, \dodoi{10.1088/0004-637X/698/2/1943}

\bibitem[{Xiao {et~al.}(2023)Xiao, Oesch, Elbaz, Bing, Nelson, Weibel, Naidu, Daddi, Bouwens, Matthee, Wuyts, Chisholm, Brammer, Dickinson, Magnelli, Leroy, van Dokkum, Schaerer, Herard-Demanche, Barrufet, Endsley, Fudamoto, G{\'{o}}mez-Guijarro, Gottumukkala, Illingworth, Labbe, Magee, Marchesini, Maseda, Qin, Reddy, Shapley, Shivaei, Shuntov, Stefanon, Whitaker, \& Wyithe}]{Xiao2023}
Xiao, M., Oesch, P., Elbaz, D., {et~al.} 2023, arXiv preprint arXiv:2309.02492.
\newblock \doarXiv{2309.02492}

\bibitem[{Xie {et~al.}(2024)Xie, {De Lucia}, Fontanot, Hirschmann, Bah{\'{e}}, Balogh, Muzzin, Vulcani, Baxter, Forrest, Wilson, Rudnick, Cooper, \& Rescigno}]{Xie2024}
Xie, L., {De Lucia}, G., Fontanot, F., {et~al.} 2024, The Astrophysical Journal Letters, 966, L2, \dodoi{10.3847/2041-8213/ad380a}

\bibitem[{Yajima {et~al.}(2021)Yajima, Abe, Khochfar, Nagamine, Inoue, Kodama, Arata, {Dalla Vecchia}, Fukushima, Hashimoto, Kashikawa, Kubo, Li, Matsuda, Mawatari, Ouchi, \& Umehata}]{Yajima2021}
Yajima, H., Abe, M., Khochfar, S., {et~al.} 2021, Monthly Notices of the Royal Astronomical Society, 509, 4037, \dodoi{10.1093/mnras/stab3092}

\bibitem[{Yung {et~al.}(2022{\natexlab{a}})Yung, Somerville, Ferguson, Finkelstein, Gardner, Dav{\'{e}}, Bagley, Popping, \& Behroozi}]{Yung2022a}
Yung, L. Y.~A., Somerville, R.~S., Ferguson, H.~C., {et~al.} 2022{\natexlab{a}}, Monthly Notices of the Royal Astronomical Society, 515, 5416, \dodoi{10.1093/mnras/stac2139}

\bibitem[{Yung {et~al.}(2022{\natexlab{b}})Yung, Somerville, Finkelstein, Behroozi, Dav{\'{e}}, Ferguson, Gardner, Popping, Malhotra, Papovich, Rhoads, Bagley, Hirschmann, \& Koekemoer}]{Yung2022b}
Yung, L. Y.~A., Somerville, R.~S., Finkelstein, S.~L., {et~al.} 2022{\natexlab{b}}, Monthly Notices of the Royal Astronomical Society, 519, 1578, \dodoi{10.1093/mnras/stac3595}

\end{thebibliography}
\bibliographystyle{aasjournal}

\end{document}